\documentclass[sort&compress,numbers]{els-mrw-mod} 

\usepackage{amsmath,amssymb,amsfonts,amsthm,makeidx,graphicx}
\usepackage[dvipsnames,table,xcdraw]{xcolor}
\usepackage{txfonts}
\usepackage{helvet}
\usepackage{feynmp-auto}
\usepackage[capitalise]{cleveref}
\usepackage{subcaption}
\usepackage[normalem]{ulem}
\newcommand{\sss}{\scriptscriptstyle}
\newcommand{\sw}{\ensuremath{s_{\sss W}}}
\newcommand{\cw}{\ensuremath{c_{\sss W}}}
\newcommand{\cBB}{\ensuremath{c_{\sss BB}}}
\newcommand{\cWW}{\ensuremath{c_{\sss WW}}}
\newcommand{\cGG}{\ensuremath{c_{\sss GG}}}
\newcommand{\cH}{\ensuremath{c_{\sss H}}}
\newcommand{\cHH}{\ensuremath{c_{\sss HH}}}
\newcommand{\cXX}{\ensuremath{c_{\sss XX}}}
\newcommand{\cQ}{\ensuremath{\mathbf{c}_{\sss Q}}}
\newcommand{\cL}{\ensuremath{\mathbf{c}_{\sss L}}}
\newcommand{\cF}{\ensuremath{\mathbf{c}_{\sss F}}}
\newcommand{\cu}{\ensuremath{\mathbf{c}_{u}}}
\newcommand{\cd}{\ensuremath{\mathbf{c}_{d}}}
\newcommand{\ce}{\ensuremath{\mathbf{c}_{e}}}
\newcommand{\kU}{\ensuremath{\mathbf{k}_{\sss U}}}
\newcommand{\kD}{\ensuremath{\mathbf{k}_{\sss D}}}
\newcommand{\kL}{\ensuremath{\mathbf{k}_{\sss L}}}
\newcommand{\kF}{\ensuremath{\mathbf{k}_{\sss F}}}
\newcommand{\ku}{\ensuremath{\mathbf{k}_{u}}}
\newcommand{\kd}{\ensuremath{\mathbf{k}_{d}}}
\newcommand{\ke}{\ensuremath{\mathbf{k}_{e}}}
\newcommand{\tcBB}{\ensuremath{\tilde{c}_{\sss BB}}}
\newcommand{\tcWW}{\ensuremath{\tilde{c}_{\sss WW}}}
\newcommand{\tcGG}{\ensuremath{\tilde{c}_{\sss GG}}}

\newcommand{\gagg}{\ensuremath{g_{ a gg }}}
\newcommand{\gaa}{\ensuremath{g_{ a \gamma\gamma }}}
\newcommand{\gaZ}{\ensuremath{g_{a\gamma{\sss  Z} }}}
\newcommand{\gZZ}{\ensuremath{g_{a{\sss Z Z }}}}
\newcommand{\gWW}{\ensuremath{g_{a {\sss W W} }}}

\newcommand{\gXXp}{\ensuremath{g_{a{\sss  XX^\prime} }}}
\newcommand{\caa}{\ensuremath{c_{ \gamma\gamma }}}
\newcommand{\caZ}{\ensuremath{c_{\gamma{\sss  Z} }}}
\newcommand{\cZZ}{\ensuremath{c_{{\sss Z Z }}}}
\newcommand{\Ceffgg}{C_{\scriptscriptstyle GG}^\text{eff} }
\newcommand{\Ceffaa}{C_{\gamma\gamma}^\text{eff} }
\newcommand{\CeffZZ}{C_{\scriptscriptstyle ZZ}^\text{eff} }
\newcommand{\CeffWW}{C_{\scriptscriptstyle WW}^\text{eff} }
\newcommand{\CeffaZ}{C_{\gamma{\scriptscriptstyle Z}}^\text{eff} }
\newcommand{\CeffZh}{C_{{\scriptscriptstyle Z}h}^\text{eff} }
\newcommand{\Ceffah}{C_{ah}^\text{eff} }
\newcommand{\Ceffff}{C_{ff}^\text{eff} }
\newcommand{\CP}{$C\kern-0.7pt P$}
\begin{document}


\chapter{Axions and Axion-like particles: collider searches}\label{chap:ALPcollider}

\author[1]{Anke Biek{\"o}tter}%
\author[2]{Ken Mimasu}%


\address[1]{\orgname{Karlsruhe Institute of Technology}, \orgdiv{Institute for Theoretical Particle Physics}, \orgaddress{Wolfgang-Gaede-Straße 1, 76131 Karlsruhe, Germany}}
\address[2]{\orgname{University of Southampton}, \orgdiv{School of Physics and Astronomy}, \orgaddress{Highfield, Southampton S017 1BJ, United Kingdom}}

\articletag{Chapter Article tagline: update of previous edition, reprint.}

\maketitle

\begin{abstract}[Abstract]
	We give an overview of collider searches for Axion-like particles~(ALPs). The intention of this review is to give a pedagogical introduction to collider phenomenology of ALPs, and provide a starting point for newcomers, including suitable references to deepen their knowledge. 
    We motivate how ALPs arise from the breaking of approximate global symmetries and describe their interactions across different scales in an effective field theory framework. We further review the dominant production and decay channels for ALPs at high-energy hadron and lepton colliders as well as indirect ways to probe their interactions via precision measurements of Standard Model processes. 
\end{abstract}

\begin{keywords}
 	Axion-like particles\sep collider physics\sep phenomenology \sep effective field theory
\end{keywords}



\section*{Objectives}
\begin{itemize}
	\item Theoretical motivation for Axion-like particles
	\item Effective field theory of Axion-like particle interactions: degrees of freedom, basis choice and renormalization group evolution
	\item Axion-like particle production and decay at hadron and lepton collider experiments
    \item Indirect and non-resonant searches for Axion-like particles
    \item Summary of current collider bounds on Axion-like particles coupled to photons
\end{itemize}

\section{Introduction}\label{sec:intro}
The last few decades of high-energy collider experiments have led to a major paradigm shift, challenging our preconceptions about new physics at the TeV scale. It is essential to ask ourselves what we can learn from the data, notably from the absence of direct evidence for physics beyond the standard model (BSM) at the Large Hadron Collider (LHC). One possibility is that new physics could be more exotic than originally thought and may be present in signatures that we have not yet considered. Alternatively, the scale of new physics, $\Lambda_{\text{BSM}}$, may be sufficiently above the electroweak (EW) scale, such that new particles are too heavy to be directly produced at the LHC. In this case, we must rely on indirect search strategies, in which precise standard model (SM) measurements may yield unexpected deviations. These can be interpreted in scenarios with heavy new states via the framework of effective field theory (EFT). 
If, instead, relatively light new particles exist, their couplings to the SM must be rather suppressed, since we have not conclusively observed them, although collecting more data may yet shed light on such scenarios. 

Perhaps a more realistic case could be that the underlying theory comprises some combination of these scenarios,
as in the case of Axion-like particles (ALPs). This scenario introduces a singlet, pseudoscalar state, $a$, with a mass, $m_a$, around or below the EW scale. The fact that it has not yet been observed suggests that its couplings to the SM should not be too large. Moreover, since it is the only addition to the particle spectrum, one may also want to motivate why it remains relatively light compared to $\Lambda_{\text{BSM}}$, where we might expect a plethora of new particles to reside. 
A simple motivation for both of these observations, which is one of the defining features of this scenario, is that the ALPs are pseudo-Nambu-Goldstone bosons (pNGBs) of an unknown, spontaneously broken, global symmetry.  First, the fact that some additional broken symmetry exists, leaving ALPs as a low-energy remnant, implies the presence of additional new physics and therefore an EFT expansion to naturally suppress ALP interactions. Second, their pNGB nature explains why they are relatively light compared to $\Lambda_{\text{BSM}}$. This is analogous to the description of pions in low-energy QCD, where role of the `ultra-violet' (UV) physics is played by the heavier bound states such as the $\rho$-meson.
In practice, the pNGB nature manifests itself in 
a shift-symmetry of the ALP field, $a\to a+\textit{c}$  where $c$ is a constant, under which the ALP interaction Lagrangian should be approximately invariant. This means that the ALP couples preferentially via \emph{derivative} interactions:
\begin{align}
    \mathcal{L}_{\text{ALP}}\supset \mathcal{L}\left(\partial_\mu a\right).
\end{align}
 The mass term $\sim m_a^2\,a^2$ explicitly breaks this symmetry, which was only taken to be approximate. However, this term (and any other explicit breaking terms such as those describing a potential for the ALP) are assumed to be sub-leading in the ALP description.

In summary, ALPs are SM singlet, spin-0 fields that are odd under parity transformations and interact primarily via derivative couplings. Altogether, this implies that the leading interactions with the SM are described by an EFT, and ALP couplings are intrinsically suppressed by a dimensionful parameter, conventionally referred to as the decay constant, $f$, 
\begin{align}
    \mathcal{L}_{\text{ALP}}\sim \mathcal{O}\left(\frac{a}{f}\right).
\end{align}
This is 
analogous to the pion decay constant, $f_{\pi}$, that suppresses pion couplings in chiral perturbation theory. This ALP decay constant is typically assumed to be related but not exactly equal to $\Lambda_{\text{BSM}}$, much like $f_{\pi}$ is to $\Lambda_{\text{QCD}}$. 
In postulating a relatively light, new particle with suppressed interactions, the ALP scenario aligns well with the current experimental status in high-energy physics. Moreover, their EFT description, on one hand, explains why ALP couplings are weak, and on the other, requires the presence of yet new BSM states at higher energies, also motivating indirect new physics searches. ALPs are therefore a timely and experimentally motivated benchmark scenario to consider in high-energy physics, and it is evident that they have grown significantly in popularity in the last decade.

Besides being experimentally motivated, ALPs are also highly motivated from a theoretical point of view. Most of us will have encountered Axions in connection with the famous solution to the strong \CP\ problem~\cite{Peccei:1977hh,Peccei:1977ur,Wilczek:1977pj,Davidson:1981zd}. 
However, global symmetries are ubiquitous in quantum field theory, and it is reasonable to expect that more may exist beyond those of the SM, and that their breaking may lead to light pNGBs. Indeed ALPs can arise in many BSM scenarios such as string compactifications~\cite{Ringwald:2012cu,Cicoli:2013ana}, supersymmetric theories~\cite{Bellazzini:2017neg}, neutrino-mass-generation mechanisms~\cite{Chikashige:1980ui,Alexandre:2020tba,Mavromatos:2020hfy}, as well as composite-Higgs realizations featuring additional pNGBs~\cite{Gripaios:2009pe}. Moreover, ALPs can be viable dark matter (DM) candidates or serve as mediators to the DM sector, see {\it e.g.}~\cite{Preskill:1982cy,Abbott:1982af,Dine:1982ah,Dror:2023fyd,Fitzpatrick:2023xks,Ghosh:2023tyz}.

Specific models may lead to relations between the properties and interactions of the ALP; for example, in typical Axion solutions to the strong \CP\ problem, the ALP mass is inversely proportional to its decay constant. However, it is useful to pursue a model-independent, bottom-up approach to cover as broad a range of scenarios as possible. We therefore treat $m_a$ 
and $f$ as independent, free parameters along with the interaction strengths of the ALP with the various SM particles, described by the ALP-EFT. 
Our only requirement is that the EFT is a valid description of the high scale physics at energies below $\Lambda_{\text{BSM}}$ and, in particular, we must have $m_a \ll \Lambda_{\text{BSM}}$.

An EFT is constructed by writing down the most general set of interactions between its degrees of freedom that respect its symmetries. The particle content of the ALP-EFT corresponds simply to the SM states plus $a$.
In this review, we are concerned with high-energy collider probes of ALPs, and so it is appropriate to consider the full, unbroken gauge symmetry group of the SM. This ensures that the underlying UV theory can reproduce the SM at low energies. The SM part of this EFT is known as the Standard Model Effective Field Theory (SMEFT) and describes gauge-invariant, effective interactions between SM states, and the ALP-EFT amounts to adding analogous terms involving the ALP.
Since we have not yet definitively pinned down the nature of EW symmetry breaking, assuming a linear realization (\emph{i.e.}, taking the Higgs as an $S\!U(2)$ doublet field) may be considered restrictive. Non-linear EW symmetry breaking can be described by the EW chiral Lagrangian~\cite{Longhitano:1980tm,Appelquist:1980vg,Herrero:1993nc} and the associated EFT that incorporates a dynamical Higgs is sometimes referred to as Higgs effective field theory, or HEFT~\cite{Alonso:2012px,Buchalla:2012qq,Buchalla:2013rka}. Although we do not discuss it any further in this review, choosing an EFT that implements a non-linear realization of EW symmetry breaking would lead to a different ALP-EFT, which is often referred to as the non-linear ALP EFT~\cite{Brivio:2017ije}.

Since ALPs can arise from a variety of UV completions, in principle there is no prior on $m_a$, nor its couplings to the SM states. Many concrete models predict more than one ALP candidate with different masses, further motivating the exploration of ALPs across multiple scales. The only certainty is that, as will be discussed in~\cref{sec:ALP-RGE}, renormalization group evolution (RGE) mixes the coefficients of the ALP EFT order-by-order in the expansion, meaning that assuming the absence of particular ALP couplings is a scale-dependent statement and one is bound to generate all interactions allowed by the underlying symmetries at some loop order. 

\begin{figure}[t!]
    \centering
    \includegraphics[width=0.75\textwidth]{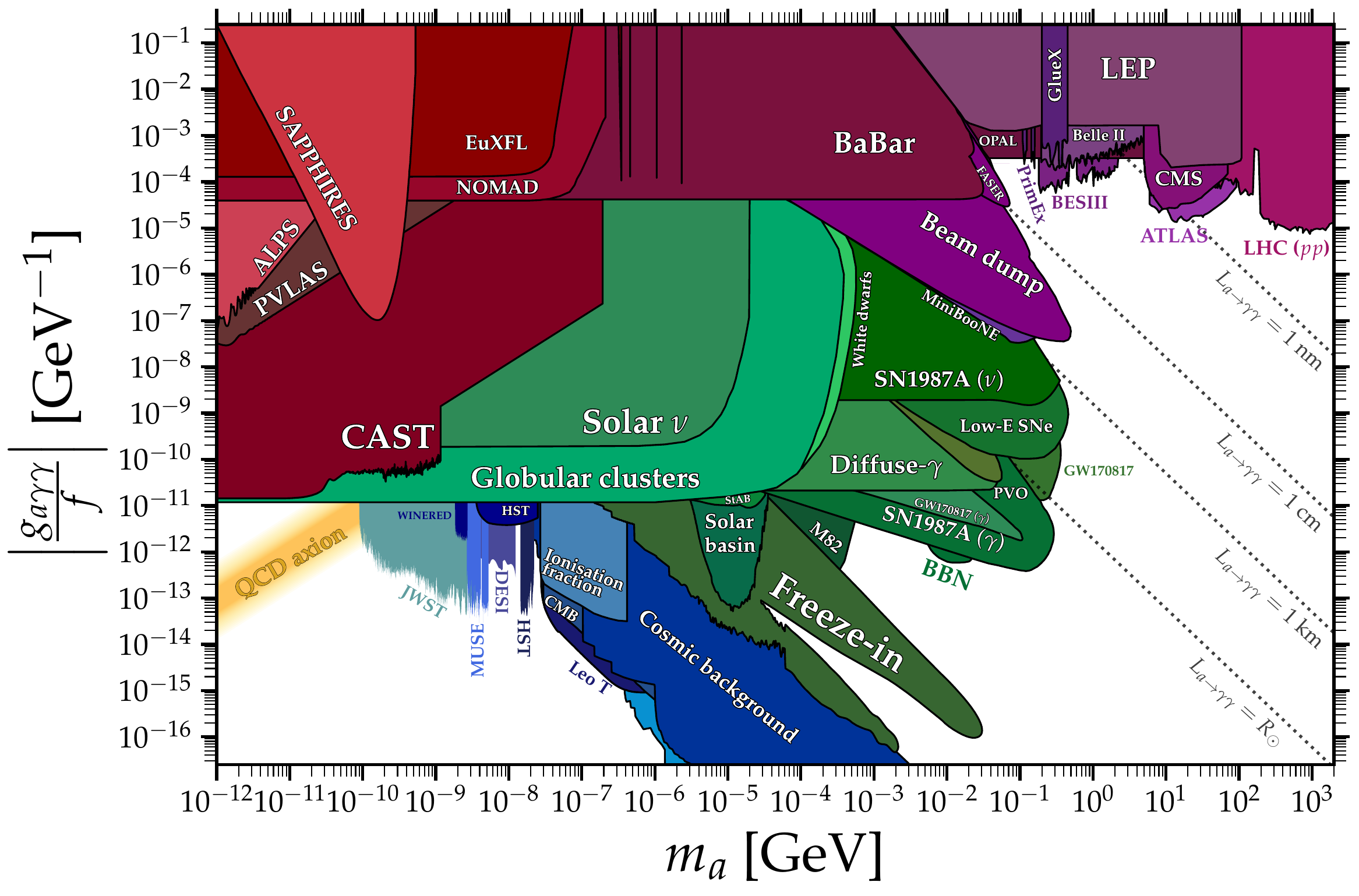}
    \caption{\label{fig:AxionLimits}
    Summary of bounds on the ALP-photon coupling adapted from Ref.~\cite{AxionLimits}; primary sources can be found in Refs.~\cite{Riordan:1987aw,Bjorken:1988as,Hearty:1989pq,Blumlein:1990ay,L3:1995nbq,NOMAD:2000usb,OPAL:2002vhf,CAST:2007jps,BaBar:2008aby,BaBar:2010eww,PrimEx:2010fvg,Ehret:2010mh,Cadamuro:2011fd,Wouters:2013hua,HESS:2013udx,Ayala:2014pea,Payez:2014xsa,Vinyoles:2015aba,Jaeckel:2015jla,Knapen:2016moh,Fermi-LAT:2016nkz,Perez:2016tcq,Dolan:2017osp,Jaeckel:2017tud,CAST:2017uph,Marsh:2017yvc,CMS:2018erd,Dobrich:2019dxc,Ng:2019gch,Aloni:2019ruo,Reynolds:2019uqt,ATLAS:2020hii,Belle-II:2020jti,Dessert:2020lil,Li:2020pcn,NA64:2020qwq,Lucente:2020whw,Depta:2020wmr,Dessert:2021bkv,Reynes:2021bpe,Foster:2021ngm,Wadekar:2021qae,Caputo:2021rux,Dolan:2021rya,Langhoff:2022bij,SAPPHIRES:2022bqg,DeRocco:2022jyq,Dolan:2022kul,Roach:2022lgo,Noordhuis:2022ljw,Caputo:2022mah,Calore:2022pks,BESIII:2022rzz,Jacobsen:2022swa,Davies:2022wvj,Hoof:2022xbe,Dessert:2022yqq,Sun:2023acy,Diamond:2023cto,Nguyen:2023czp,TOTEM:2023ewz,Capozzi:2023ffu,Todarello:2023hdk,Wang:2023imi,Janish:2023kvi,Liu:2023nct,Carenza:2023qxh,Diamond:2023scc,Muller:2023vjm,Pybus:2023yex,ATLAS:2023zfc,MAGIC:2024arq,FASER:2024bbl,Candon:2024eah,CAST:2024eil,Ning:2024eky,Ruz:2024gkl,BESIII:2024hdv,Manzari:2024jns,Yin:2024lla,Todarello:2024qci,Porras-Bedmar:2024uql,SAPPHIRES:2024xmw,Fiorillo:2025yzf}.
}
\end{figure}

Our review will focus on a particular ALP mass range in which colliders can offer complementary sensitivities to existing bounds from other sources. 
Ref.~\cite{AxionLimits} is an extremely useful resource for ALP-hunters which compiles many existing bounds on various ALP couplings. 
We emphasize, however, that many of these searches probe individual ALP couplings and their sensitivity may vary in the case where more than one type of ALP coupling is present. Moreover, they may also only apply over a range of ALP couplings, and become irrelevant when the ALP is too strongly coupled, or too long-lived for example. ~\Cref{fig:AxionLimits} shows a typical example of a collection of limits on the ALP-photon coupling, $\gaa$, defined in~\cref{eq:LEW_mass}, over a wide mass range, and assuming only this coupling is present.
Along with the bounds from various sources, the figure indicates iso-contours of constant decay length of the ALP, assuming only the $a\to\gamma\gamma$ decay channel is possible. In the lower left, the ``QCD Axion'' band indicates the region in which the simplest Axion models predict the coupling to lie, according to the aforementioned inverse relation between $m_a$ and $f$.

Broadly speaking, we can group experimental searches for ALPs into three categories.
The first concerns the well known bounds on ALPs arising from astrophysics and cosmology. These tend to target the ALP couplings to photons, electrons and nucleons, and come from a variety of sources such as axion helioscopes like CAST~\cite{CAST:2007jps,CAST:2017uph,CAST:2024eil} searching for an ALP flux from our sun ($m_a\lesssim$ eV), modifications of the stellar populations in globular clusters~\cite{Ayala:2014pea,Dolan:2022kul} ($m_a\lesssim$ keV), observations of supernova 1987a ($m_a\lesssim$ MeV) ~\cite{Payez:2014xsa,Jaeckel:2017tud,Lucente:2020whw,Caputo:2021rux,Caputo:2022mah,Hoof:2022xbe,Diamond:2023scc,Muller:2023vjm,Manzari:2024jns}, and many more~\cite{Cadamuro:2011fd,Wouters:2013hua,HESS:2013udx,Vinyoles:2015aba,Fermi-LAT:2016nkz,Perez:2016tcq,Marsh:2017yvc,Ng:2019gch,Reynolds:2019uqt,Dessert:2020lil,Li:2020pcn,Dessert:2021bkv,Reynes:2021bpe,Foster:2021ngm,Wadekar:2021qae,Dolan:2021rya,Langhoff:2022bij,DeRocco:2022jyq,Roach:2022lgo,Noordhuis:2022ljw,Calore:2022pks,Jacobsen:2022swa,Davies:2022wvj,Dessert:2022yqq,Sun:2023acy,Diamond:2023cto,Nguyen:2023czp,Todarello:2023hdk,Wang:2023imi,Janish:2023kvi,Liu:2023nct,Carenza:2023qxh,MAGIC:2024arq,Candon:2024eah,Ning:2024eky,Ruz:2024gkl,Yin:2024lla,Todarello:2024qci,Porras-Bedmar:2024uql,Fiorillo:2025yzf}. The $m_a$ ranges next to each set of bounds indicate that each astrophysical phenomenon used to set a constraint has a characteristic temperature that limits the ALP mass range to which it is sensitive and that these types of bounds only generally apply up to $m_a\sim 100$~MeV, as seen in ~\cref{fig:AxionLimits}. In some cases, the bounds are sensitive to a characteristic timescale of the ALP decay, \emph{i.e.}, they probe ALPs that have a specific lifetime. For example, Big Bang Nucleosynthesis bounds~\cite{Depta:2020wmr} are sensitive to lifetimes of order 1 second. Such bounds will depend more strongly on a combination of $m_a$ and the set of ALP couplings to SM states. Notably, to the left of the point at which the ``Cosmic background'' limits apply in~
\cref{fig:AxionLimits}, the ALP lifetime is longer than the age of the universe, and it can act as cold DM.

A second category of ALP bounds come from low-energy experiments such as ``Light shining through walls''~\cite{Ehret:2010mh,SAPPHIRES:2022bqg,SAPPHIRES:2024xmw,NOMAD:2000usb} and beam-dump experiments. These experiments look to experimentally produce a flux of ALPs that travel a finite distance through some shielding material (the ``wall'') before decaying in an instrumented region. Such experiments also probe ALPs of specific lifetimes given by the baseline of the experiment. The ALP mass to which the experiment is sensitive then depends on the production mechanism, and current bounds extend out to the hundreds of MeV range, but generally probe a complementary region with stronger ALP couplings.

The third category, from the 100~MeV to GeV regime and beyond, is the realm of collider physics, understandably so given that these are the highest-energy experiments available to us. Examples of such bounds can be seen in the upper right part of~\cref{fig:AxionLimits} and will be discussed in more detail, along with beam dump experiments in~\cref{sec:bounds}. Here one can further split this category into flavor and high-energy collider physics. The former involves searching for ALPs through the decays of mesons such as pions, kaons, and $B$-mesons and efficiently probes the region up to about $m_a\sim5$~GeV, providing the best bounds in the 100~MeV-5~GeV range. We will discuss these bounds in more detail in~\cref{sec:prod_in_decays}. 

Above the mass of the $B$-mesons, ALPs cannot be produced via these decays and must therefore be produced through the decay of heavier particles such as the Higgs and $Z$ bosons or directly through hard scatterings at collider experiments. Independently of $m_a$, colliders enable the first direct probes of ALP couplings to the heavy states in the SM, since these can only be produced at high energy machines. Bounds on such couplings from colliders would be complimentary across the whole mass range. However, the fact that the ALP couplings mix under RGE implies that, at some level, these couplings will be constrained indirectly by non-collider bounds, which are generally much more sensitive and yield limits that are well beyond collider sensitivity, even when arising at a loop-suppressed level. Given the focus on collider phenomenology of this review, we will not consider ALPs with masses below the hundreds of MeV range.  

The rest of the review is organized as follows. We begin in~\cref{sec:toy} with a pedagogical example of a complex scalar field theory giving rise to a pNGB.
In~\cref{sec:interactions}, we review the generic interactions of the ALP in the framework of an EFT at the EW scale assuming a linearly-realized EW symmetry breaking. In~\cref{sec:production}, we discuss the main production modes of ALPs at colliders, and in~\cref{sec:decay} the possible decay modes of the ALP into SM states. In~\cref{sec:indirect}, we focus on indirect constraints on ALPs through non-resonant production and ALP-SMEFT interference. In~\cref{sec:bounds} we summarize some key constraints from colliders on ALPs, taking the ALP-photon coupling as an example.  Finally, in~\cref{sec:conclusions} we briefly conclude  and provide some additional references to reviews focusing on complementary aspects of Axions and ALPs.

\section{Toy example}\label{sec:toy}
In the following, we highlight how pNGBs arise from spontaneously broken approximate global symmetries. We show how a $U(1)$ symmetry results in having a shift-symmetric angular mode, characterized by derivative couplings to other particles, and that these couplings are suppressed by powers of the vacuum-expectation-value (vev) of the complex field it originates from. 
We consider a complex scalar field, $\varphi$, described by the following Lagrangian density:
\begin{align}
\begin{split} 
    \mathcal{L}_\varphi &= |\partial_\mu\varphi|^2 - V(\varphi),\quad
    V(\varphi)=-\mu^2|\varphi|^2+\lambda|\varphi|^4.
\end{split}
\end{align}
$\mathcal{L}_\varphi$ has a global $U(1)$ symmetry associated with a rephasing, 
$\varphi\to e^{i\theta}\varphi$.
When $\mu^2>0$, the potential $V(\varphi)$ leads to a non-zero vev for the modulus of $\varphi$,
\begin{align}
    \langle\varphi\rangle = \sqrt{\frac{\mu^2}{2\lambda}}\equiv \frac{f}{\sqrt{2}},
\end{align}
which spontaneously breaks the global $U(1)$ symmetry.
We can express the complex field $\varphi$ in polar coordinates in terms of two real scalar fields, $r(x)$ and $a(x)$, 
\begin{align}
    \varphi(x) = \frac{r(x)+f}{\sqrt{2}}\exp\left(i\frac{a(x)}{f}\right),
\end{align}
where $r(x)$ and $a(x)$ are defined such that they do not acquire a vev. The $U(1)$ transformation leads to a shift: $a(x)\to a(x)+\mathrm{constant}$.

The scalar potential is independent of $a(x)$ but contains a mass, $m_r^2=2\lambda f^2$,  as well as trilinear and quartic self-interactions for $r(x)$ 
\begin{align}
     V(\varphi) =  V(r) = \lambda f^2 r^2 + \lambda f r^3 + \frac{\lambda}{4} r^4 + \mathrm{constant}.
\end{align}
The kinetic term, instead, depends on both fields and takes the form
\begin{align}
\label{eq:LKin}
    \mathcal{L}^{\mathrm{kin.}}_\varphi=
    \frac{1}{2}\left|
    \partial_\mu r\,e^{i\frac{a}{f}}+ i \big(r+f\big) \,\frac{\partial_\mu a}{f} \,e^{i\frac{a}{f}}\right|^2
    =\frac{1}{2}\partial_\mu r\,\partial^\mu r + \frac{1}{2}\left(1 + 2\frac{r}{f}+\frac{r^2}{f^2}\right)\partial_\mu a\,\partial^\mu a.
\end{align}
In addition to the kinetic terms for the two fields, it contains interactions  involving derivatives of $a(x)$ that are manifestly shift-invariant. 
After the spontaneous breaking of the global $U(1)$ symmetry, our theory describes a massive Higgs boson and a massless Nambu-Goldstone boson.

Generating a mass term for $a(x)$ requires an explicit breaking of the global symmetry. This can be realized, \emph{e.g.}, by adding to the potential:
\begin{align}
\label{eq:Lbreak}
    V_\mathrm{break} &= \,-\frac{m_a^2}{4}\left(\varphi^2+(\varphi^\ast)^2\right)=-\frac{m_a^2}{4}(r+f)^2\cos\left(\frac{2\,a}{f}\right).
\end{align}
The polar representation on the right-hand side of~\cref{eq:Lbreak} shows how the $U(1)$ breaking term generates a potential for $a(x)$ along with some additional interaction terms with $r(x)$. The vev of the modulus of $\varphi$ is modified
\begin{align}
    \langle\varphi\rangle=\frac{1}{2}\sqrt{\frac{2\mu^2+m_a^2}{\lambda}} \equiv \frac{f}{\sqrt{2}},
\end{align}
and the new potential is still minimized at $\langle a\rangle=0$. Expanding~\cref{eq:Lbreak} around $a=0$ gives the corrections to the potential,
\begin{align}
    \label{eq:Vbreak_expanded}
    V(\varphi)+V_\mathrm{break}&\simeq V(r)+\frac{1}{2}m_a^2a^2\left(
    1+2\frac{r}{f} + \frac{r^2}{f^2} 
    \right) + \mathcal{O}(a^4),
\end{align}
showing that $a$ is now a pNGB, having gained a mass, $m_a$, from the explicit breaking along with some additional interactions, all of which break its shift symmetry. Note that the potential in~\cref{eq:Lbreak} retains a discrete version of the shift symmetry $a\to a+n\pi f$, for any integer, $n$.\footnote{Note that the same exercise can be done using the Cartesian form, $\varphi = (h(x)+i\chi(x))/\sqrt{2}$, leading to the same conclusions. In this field basis, the shift symmetry is not manifest but all interactions remain at dimension four, unlike in~\cref{eq:LKin,eq:Vbreak_expanded}.} 

To make contact with the general description of the ALP-EFT in the previous section, from this toy example we would integrate out the radial mode, $r$, and identify $m_r$ with $\Lambda_\text{BSM}$ to obtain the low-energy EFT.

\section{Interactions of an Axion-like particle}\label{sec:interactions}

As discussed in the introduction, the ALP interactions are described by an EFT whose low-energy degrees of freedom are the SM particles plus the ALP, that respects the symmetries of the SM as well as an approximate shift-symmetry for the ALP.
Up to mass dimension six, the ALP effective Lagrangian that fulfills these conditions is given by
\begin{align}
\label{eq:ALP_lag}
\begin{split}
   \mathcal{L}^{D\le 6} = & \frac{1}{2}\,(\partial_\mu a ) (\partial^\mu a ) - \frac{m_a^2}{2}\,a^2 + 
   \,\cGG\,\frac{a}{f}\,\frac{\alpha_s}{4\pi}\,G_{\mu\nu}^A\,\widetilde G^{\mu\nu\,A} + \cWW\,\frac{\alpha_L}{4\pi}\,\frac{a}{f}\,W_{\mu\nu}^I\,\widetilde W^{\mu\nu\,I} + \cBB\,\frac{\alpha_Y}{4\pi}\,\frac{a}{f}\,B_{\mu\nu}\,\widetilde B^{\mu\nu} 
   \\
   & + \frac{\partial^\mu a}{f}\,\sum_F \bar F\, \cF \,\gamma_\mu\, F 
   + \frac{c_H}{f}\, \partial^\mu a\,(H^\dagger i\overleftrightarrow
D_\mu H)
   + \frac{\cHH}{f^2}\,(\partial^\mu a)(\partial_\mu a)\,H^\dagger H\,,
\end{split}
\end{align}
where $\cF$ ($F=Q,L,u,d,e$) are $3\times3$ hermitian (or symmetric in the \CP-conserving case) matrices, one for each of the fermionic representations of the SM, and $c_{ii}$ ($i=G,W,B,H$) are real couplings~\cite{Georgi:1986df}. 
Factors of $\alpha_i/(4 \pi)$ have been pulled out from the interaction terms with the SM gauge fields.
Note that by doing so, we have implicitly assumed that the bosonic ALP couplings are loop suppressed, while the fermionic ALP couplings arise at tree level. This will become relevant when we discuss alternative forms of the ALP Lagrangian below. 

The ALP mass term is the only term that explicitly breaks the shift symmetry, $a \to a + c$. The gauge boson couplings do not involve a derivative of the ALP field and therefore do not obviously respect the shift symmetry. 
However, performing such a shift on the gauge boson operators
\begin{align}
    \frac{a}{f}F^i_{\mu\nu}\widetilde{F}_i^{\mu\nu}&\to
    \frac{a+c}{f}F^i_{\mu\nu}\widetilde{F}_i^{\mu\nu},    
\end{align}
has the effect of shifting the topological $\theta$-term of the corresponding gauge group.
Such a term can be written as total derivative
\begin{align}
\label{eq:total_derivative}
F^i_{\mu\nu}\widetilde{F}_i^{\mu\nu} = \partial^\mu K_\mu,\quad K_\mu = 2\epsilon_{\mu\nu\rho\sigma} \left(A^\nu_i\partial^\rho A_i^\sigma+\frac{ig}{3}f_{ijk}A^\nu_i A^\rho_j A^\sigma_k\right),
\end{align}
where $i$ indicates the adjoint index of an $S\!U(N)$ gauge group with structure constants, $f_{ijk}$, which does not affect the action at the perturbative level. At the non-perturbative-level, only $S\!U(3)_C$ $\theta$-term turns out to be physical,\footnote{For non-Abelian gauge theories the total derivative in ~\cref{eq:total_derivative} is physical due to the presence of spatially extended instanton solutions of the field equations, such that $\int d^4x \,\partial^\mu K_\mu$ does not vanish when evaluated at spatial infinity~\cite{Polyakov:1975rs,Belavin:1975fg,tHooft:1976rip,tHooft:1976snw,Callan:1976je,Jackiw:1976pf}. In terms of the gauge groups of the SM, the term for $U(1)_{Y}$ is therefore unphysical. The $S\!U(2)_{L}$ term also turns out to be unphysical because it can be removed by the anomaly-induced effect of performing a $B+L$ transformation on the left-handed fermion fields (see, \emph{e.g.}, ~\cite{Anselm:1993uj,FileviezPerez:2014xju}). This leaves only the $S\!U(3)_C$ term, whose  physicality is precisely the origin of the strong \CP\ problem.
} meaning that the ALP-gluon coupling non-perturbatively breaks the shift symmetry. In this review, we will only concern ourselves with perturbative observables, and hence where ALP-gauge couplings are shift-symmetry preserving.

The dimension-five operator  $\mathcal{O}_H = \partial^\mu a\,(H^\dagger i\overleftrightarrow
D_\mu H)$ is redundant as it can be rewritten in terms of the ones in~\cref{eq:ALP_lag} by means of field redefinitions~\cite{Bauer:2020jbp}, which -- despite leading to different-looking Lagrangians --  always preserve $S$-matrix elements.
This can be seen by redefining the fields in the SM Lagrangian by 
\begin{align}
\label{eq:redef}
H &\to H \exp \left[i \, x_{\sss H} \frac{a}{f} \right], 
& 
F &\to F \exp \left[ i \, \mathbf{x}_{\sss F} \frac{a}{f} \right] \, ,
\end{align}
with real rotation parameters $x_{\sss H}$ and $\mathbf{x}_{\sss F}$, and where the latter is in general a matrix in flavor space. A key feature of these field redefinitions is that, being chiral transformations, the fermionic path integral measure transforms non-trivially~\cite{Fujikawa:1979ay}, inducing shifts of the ALP-gauge--boson couplings, $\cXX$, at order $\alpha_i/4\pi$.
The selection of a specific basis is a choice; here we choose to work in a basis without the operator~$\mathcal{O}_H$. 
Note however, that even in a basis including $\mathcal{O}_H$, the predicted $a$-$Z$-$h$ and $a$-$Z$-$h$-$h$ vertices do not contribute to amplitudes with the $Z$-boson on-shell that one might have expected to mediate, \emph{e.g.}, $h\to Za$ or $a\to Zh$ ~\cite{Bauer:2016zfj,Bauer:2020jbp}.
The term parametrized by $\cHH$ is the only shift-symmetric, dimension-six interaction of the ALP and describes its leading interaction with the SM Higgs boson.
Tree-level couplings of the ALP to a SM Higgs boson and a $Z$ boson, as needed for the decay $h \to Z a$, are first induced at mass dimension seven.

Even after removing the operator $\mathcal{O}_H$, there are some remaining redundancies in the ALP Lagrangian in~\cref{eq:ALP_lag}. 
These can be removed by ALP-dependent field redefinitions of the type shown in~\cref{eq:redef}, with the $x_i$ assigned according to the quantum numbers of each state under the accidental $U(1)$ symmetries of the SM, namely baryon ($B$) and lepton ($L$) number. These redefinitions imply additional constraints on the elements of $\cF$.\footnote{The redundancy of $\mathcal{O}_H$ is analogously connected to the remaining $U(1)_Y$ hypercharge global symmetry of the SM. Redefining the fields setting $x_i$ proportional to hypercharge generates a relation purely between $\mathcal{O}_H$ and the fermion couplings, as the global symmetry is not anomalous, such that no gauge boson operators are involved.}
One can show that the ALP couplings to the baryonic and leptonic currents can be written in terms of bosonic operators
\begin{equation}
\begin{split}
\frac{\partial_\mu a}{f} J_B^\mu &= \text{Tr} \left[ \frac{\mathcal{O}_Q + 
\mathcal{O}_u + \mathcal{O}_d}{3} \right]
= \frac{N_f}{16 \pi^2 } \left( g^2 \mathcal{O}_{WW}- g^{\prime 2} \mathcal{O}_{BB} \right) \, , 
\\
\frac{\partial_\mu a}{f} J_{L_i}^\mu &= \left[\mathcal{O}_L + \mathcal{O}_e \right]^{ii}
= \frac{1}{32 \pi^2} \left( g^2 \mathcal{O}_{WW}- g^{\prime 2} \mathcal{O}_{BB} \right) 
\, ,
\end{split}
\label{eq:relation_B_L}
\end{equation}
where the $\mathcal{O}_x$ are the operators corresponding to the Wilson coefficients $c_x$ in~\cref{eq:ALP_lag}.
\Cref{eq:relation_B_L} gives us one relation between the quark and gauge couplings related to~$B$ conservation, along with one relation per lepton species, corresponding to the conservation of $L_i$ with $i= e, \mu, \tau$.
In total, we thus find one redundancy for the quark couplings and three redundancies for the diagonal lepton couplings for $N_f=3$ fermion generations.
A very nice and explicit summary of field redefinitions and operator basis reduction, is given in appendix~B of~\cite{Bonilla:2021ufe}.

In~\cref{tab:param_counting}, we list the number of independent parameters of the dimension-five ALP Lagrangian for one and three fermion families. In total, seven (46) parameters of the ALP Lagrangian are independent for $N_f=1$ ($N_f=3$) fermion families. If we make the additional assumption that the ALP Lagrangian is \CP-even, the number of independent parameters reduces to 30 for three fermion generations.
In the rest of the review, we focus on \CP-even ALP couplings, i.e.\ we consider $\cF$ to have real entries only. For \CP-violating ALP couplings, we refer to the review in~\cite{DiLuzio:2023lmd}. 
For ALP couplings up to higher mass dimensions, the number of free parameters in the ALP Lagrangian has been determined up to mass dimension 15 using Hilbert series techniques~\cite{Grojean:2023tsd}. An operator basis has been constructed for the ALP-EFT above and below the EW scale up to (and including) mass dimension eight~\cite{Song:2023lxf,Grojean:2023tsd}. 
\begin{table}[t]
    \centering
    \begin{tabular}{c|cccc}
     parameter & conditions & $N_f=1$ & $N_f=3$ (\CP-conserving) & $N_f=3$ \\
     \hline
      $m_a$   && 1 &1 & 1 \\
      $c_i/f$   & $\cGG, \, \cWW , \, \cWW, \, \cH \in \mathcal{R}$ & 4 & 4 & 4 \\
      $\cF/f$ &  $N_f \times N_f$ symmetric/hermitian matrices & $5\times 1$ & $5 \times 6$ &  $5 \times 9$\\
      \hline
	  & $Y$ rotations & $-1$ & $-1$ & $-1$ \\      
      & $L_i$ rotations & $-1$ & $-3$ & $-3$ \\
      & $(B-L)$ rotation & $-1$ & $-1$& $-1$ \\
      \hline
      & & 7 & 30 & 46
    \end{tabular}
    \caption{Number of free parameters of the dimension-five ALP Lagrangian for one or three fermion families.}
    \label{tab:param_counting}
\end{table}

A commonly used alternative form of the Lagrangian in~\cref{eq:ALP_lag} writes the ALP Lagrangian as that of a pseudoscalar. In this basis, all of the ALP-fermion couplings are chirality-flipping:
\begin{align}
\label{eq:ALP_lag_alt}
\begin{split}
   \mathcal{L}^{\prime\,D\le 6} = & \frac{1}{2}\,(\partial_\mu a ) (\partial^\mu a ) - \frac{m_a^2}{2}\,a^2 + 
   \,\tcGG\,\frac{a}{f}\,\frac{\alpha_s}{4\pi}\,G_{\mu\nu}^A\,\widetilde G^{\mu\nu\,A} 
   + \tcWW\,\frac{\alpha_L}{4\pi}\,\frac{a}{f}\,W_{\mu\nu}^I\,\widetilde W^{\mu\nu\,I} 
   + \tcBB\,\frac{\alpha_Y}{4\pi}\,\frac{a}{f}\,B_{\mu\nu}\,\widetilde B^{\mu\nu} 
   \\
   &  -\frac{a}{f} \, \left( \bar{Q} H \mathbf{\tilde{Y}}_d d_R  
   + \bar{Q} \tilde{H} \mathbf{\tilde{Y}}_u u_R 
   + \bar{L} H\mathbf{\tilde{Y}}_e e_R 
   + \text{h.c. } \right)
   +\frac{a^2}{2f^2}\, \left( \bar{Q} \phi \mathbf{Y}^\prime_d d_R  
   + \bar{Q} \tilde{\phi} \mathbf{Y}^\prime_u u_R 
   + \bar{L} \phi \mathbf{Y}^\prime_e e_R 
   + \text{h.c. } \right)
   + \frac{\cHH}{f^2}\,(\partial^\mu a)(\partial_\mu a)\,H^\dagger H\,.
\end{split}
\end{align}
One can use field redefinitions of the type given in~\cref{eq:redef} to move from the derivative basis to the chirality-flipping basis. Since there are more free parameters in the fermion sector of~\cref{eq:ALP_lag_alt}, additional requirements on the Lagrangian parameters need to be applied to ensure that it corresponds to a shift-symmetric ALP. This basis change shifts the ALP-gauge--boson couplings by combinations of traces of the original fermion couplings, $\mathrm{Tr}[\cF]$.
Specifically, the gauge boson couplings in the chirality-flipping basis are related to the derivative basis Wilson coefficients as
\begin{align}
    \tcGG &= \cGG + \frac{1}{2} \text{Tr}\left[ \cu + \cd - 2 \cQ \right] \, ,
    & 
    \tcWW &= \cWW - \frac{1}{2} \text{Tr}\left[ 3 \cQ +\cL \right]\, ,
    & 
    \tcBB &= \cBB + \text{Tr}\left[ \frac{4}{3} \cu + \frac{1}{3} \cd - \frac{1}{6}\cQ+ \ce - \frac{1}{2} \cL \right]
    \, ,
\label{eq:relation_derivative_pseudoscalar}
\end{align}
and the Yukawa-like coupling matrices $\mathbf{\tilde{Y}}_x$ and $\mathbf{Y}^\prime_x$ are given by 
\begin{align}
\begin{aligned}
\mathbf{\tilde{Y}}_d &=i \left(  \mathbf{Y}_d \cd - \cQ \mathbf{Y}_d \right) \, ,
& 
\mathbf{\tilde{Y}}_u &= i\left(  \mathbf{Y}_u \cu - \cQ \mathbf{Y}_u \right) \, ,
& 
\mathbf{\tilde{Y}}_e &=i \left(  \mathbf{Y}_e \ce - \cL \mathbf{Y}_e \right) \, , 
\\
\mathbf{Y}^\prime_d &= \left(  \mathbf{Y}_d \cd^2 - 2 \cQ \mathbf{Y}_d \cd + \cQ^2 \mathbf{Y}_d \right) \, ,
& 
\mathbf{Y}^\prime_u &= \left(  \mathbf{Y}_u \cu^2 - 2 \cQ \mathbf{Y}_{u} \cu +  \cQ^2 \mathbf{Y}_u \right) \, ,
& 
\mathbf{Y}^\prime_e &= \left(  \mathbf{Y}_e \ce^2 - 2 \cL \mathbf{Y}_{e} \ce +  \cL^2 \mathbf{Y}_e \right) \, .
\end{aligned}
\end{align}
The derivative ALP couplings to fermions can be entirely traded for Yukawa-like operators that, after EW symmetry breaking~(EWSB), lead to \CP-odd Yukawa interactions for the ALP as well as four-point $h$-$a$-$f_L$-$\bar{f}_R$ couplings. Crucially, these couplings are weighted by the corresponding Yukawa matrices of each fermion species and ultimately lead to interactions that are proportional to each fermion mass, as shown in the next section. 
The fermionic contributions to the ALP-boson couplings in~\cref{eq:relation_derivative_pseudoscalar} arise at one-loop order. As a result, the two bases are only equivalent up to one-loop effects in the ALP-fermion couplings.
Finally, we note the appearance of $a^2$ interaction terms at dimension six which must be included at this order to maintain the equivalence of the two operator bases~\cite{Bauer:2023czj}.

\subsection{The ALP Lagrangian after electroweak symmetry breaking}
After EWSB, the ALP couplings to the SM mass eigenstates are
\begin{align}
\label{eq:LEW_mass}
\begin{split}
\mathcal{L}^{D\le 6} (\mu_{\mathrm{EW}}) &=
\frac{1}{2}\,(\partial_\mu a ) (\partial^\mu a ) - \frac{m_a^2}{2}\,a^2
+ \frac{a}{f}\left(
\gagg \, G_{\mu\nu}^A\,\widetilde G^{\mu\nu\,A}
+ \gaa \,F_{\mu\nu}\,\widetilde F^{\mu\nu}
+\gaZ\,F_{\mu\nu}\,\widetilde Z^{\mu\nu}
+\gZZ\,Z_{\mu\nu}\,\widetilde Z^{\mu\nu}
+\gWW \,W^+_{\mu\nu}\,\widetilde W^{\mu\nu}_{-}
\right) \\
&+\frac{2e}{\sw}\frac{a}{f}
\gWW
\left(
W^+_\mu W^-_\nu\left(
\sw\widetilde{F}^{\mu\nu} + \cw\widetilde{Z}^{\mu\nu}
\right)+
\left(\sw A_\mu + \cw Z_\mu\right)\left(
W^+_\nu \widetilde{W}_{-}^{\mu\nu}-
W^-_\nu \widetilde{W}_{+}^{\mu\nu}
\right)
\right)\\
& + \mathcal{L}_{\mathrm{ferm}}(\mu_{\mathrm{EW}}) 
+ \frac{\cHH}{2 f^2} (\partial^\mu a)(\partial_\mu a) (h^2 + 2 v h),
\end{split}
\end{align}
where $e$ is the electromagnetic coupling constant and $\sw$ and $\cw$ denote the sine and cosine of the Weinberg angle, respectively.\footnote{Note that we have redefined the ALP field as $a \to a \, \left(1 - \cHH v^2/(2f^2)\right)$ to recover a canonical form for the ALP kinetic term. }

The mass-mixing in the neutral gauge sector leads to the following relations between the mass-basis couplings and the Wilson coefficients of~\cref{eq:ALP_lag}
\begin{align}
\label{eq:coup_gauge}
\begin{aligned}
    \gagg &= \frac{\alpha_s}{4 \pi} \cGG , & 
    \gaa &= \frac{\alpha}{4\pi} \caa = \frac{\alpha}{4\pi}(\cBB + \cWW),   &   
    \gZZ &= \frac{\alpha}{4\pi \sw^2 \cw^2} \cZZ = 
    \frac{\alpha}{4\pi \sw^2 \cw^2} ( \cw^4 \cWW + \sw^4  \cBB) ,  
    \\ 
    & & 
    \gaZ &=\frac{\alpha}{2\pi \sw \cw} \caZ
    =\frac{\alpha}{2\pi \sw \cw}(\cw^2 \cWW - \sw^2 \cBB) , & 
    \gWW &=\frac{\alpha}{2\pi \sw^2} \cWW  \, ,
\end{aligned}
\end{align}
where $\alpha$ is the electromagnetic fine structure constant. 
The two Wilson coefficients, $\cBB$ and $\cWW$, lead to four couplings between the ALP and the EW gauge bosons. It is therefore not consistent in general to consider each of these phenomenological couplings separately. At most one of these four couplings can be switched off by imposing a relation between the two Wilson coefficients. An interesting physical limit is the photophobic scenario, where $\cBB=-\cWW$ and the ALP does not couple to photons. It was shown in~\cite{Craig:2018kne} that this limit is well-defined from a model-building point of view, in that it reflects a spurious shift symmetry and is therefore stable under loop corrections up to shift-symmetry breaking effects $\propto m_a^2$. Since $\gaa$ tends to be the most constrained by non-collider data, particularly for masses below the GeV scale, this scenario represents a case in which collider probes may offer complementary information.

The ALP-fermion interactions in the fermion mass basis read
\begin{align}
\label{eq:ferm_lag_EWSB}
\mathcal{L}_{\mathrm{ferm}}(\mu_{EW}) = \frac{\partial_\mu a}{f}
\Bigg[
&\bar{u}_{\sss L} \,\kU \gamma_\mu\, u_{\sss L} 
+ \bar{u}_{\sss R}\, \ku  \gamma_\mu \,u_{\sss R}
+ \bar{d}_{\sss L} \,\kD  \gamma_\mu \,d_{\sss L}
+ \bar{d}_{\sss R}\, \kd  \gamma_\mu \,d_{\sss R}
+ \bar{\nu}_{\sss L} \,\kL  \gamma_\mu \,\nu_{\sss L}
+ \bar{e}_{\sss L} \,\kL  \gamma_\mu \,e_{\sss L}
+ \bar{e}_{\sss R} \,\ke  \gamma_\mu \,e_{\sss R}
\Bigg] \, ,
\end{align}
where the ALP-fermion couplings $\kF$ are related to the matrices $\cF$ in~\cref{eq:ALP_lag} by the unitary rotations that diagonalize the SM Yukawa 
matrices,\footnote{ The matrices $\mathbf{U}_{f, L/R}$ are defined by $\mathbf{U}_{f,L}^\dagger \mathbf{Y}_f \mathbf{U}_{f,R} = \mathbf{Y}_{f,\mathrm{diag}}$.}
\emph{e.g.}\  $\kU = \mathbf{U}_{u,L}^\dagger \cQ \mathbf{U}_{u,L}$. 
The matrices $\kU$ and $\kD$ are not independent as they are related by the CKM matrix $\mathbf{V}_\mathrm{CKM} = \mathbf{U}_{u,L}^\dagger \mathbf{U}_{d,L}$ as
\begin{align}
\kD & = \mathbf{V}_\mathrm{CKM}^\dagger \kU \mathbf{V}_\mathrm{CKM} \, .
\end{align}
At this point it is useful to briefly comment on the potential flavor structure of ALP interactions. In the SM after EWSB, the individual flavor rotation matrices, $\mathbf{U}_{f,L/R}$, only ever appear in charged-current interactions, and in the combination corresponding to $\mathbf{V}_\mathrm{CKM}$. This is a consequence of the $U(3)^5$ flavor symmetry of the SM that is only broken by the Yukawa interactions.
Non-flavor-universal ALP-fermion interactions also break $U(3)^5$ and hence lead to a dependence on the individual rotation matrices. In this case, they still only appear in the specific $\mathbf{U}^\dagger \cF\mathbf{U}$ combinations and remain unmeasurable. Indeed, the choice of flavor basis before EWSB is arbitrary, \emph{i.e.} what one defines to be first, second and third generation. Typically, one works in one of the so-called \emph{up-} or 
\emph{down-aligned} bases, where the freedom to flavor-rotate the five fermionic SM representations is used to diagonalize the Yukawa matrices for the leptons along with one of \emph{either} the up- or down-type quarks, respectively. After EWSB, the remaining freedom to rotate the other left-handed quark component is used to diagonalize its mass matrix. In practice, these basis choices correspond to setting all but one of the left-handed quark flavor rotation matrices to the identity and the remaining rotation matrix to the CKM matrix. 
\begin{align}
\mathbf{U}_{f,R}= \mathbb{I}\,\,\,\,(f=u,d,e)\,,\,\,\,   \mathbf{U}_{e,L}  = \mathbb{I}\,,\quad
   \begin{cases}
\mathbf{U}_{u,L}= \mathbb{I}\,, \,\,\,
   \mathbf{U}_{d,L}=\mathbf{V}_{\mathrm{CKM}} & (\text{up-aligned})\\
   \mathbf{U}_{d,L}= \mathbb{I}\,, \,\,\,
    \mathbf{U}_{u,L}=\mathbf{V}_{\mathrm{CKM}}^\dagger& (\text{down-aligned})  
   \end{cases}
\end{align}
This means that, even though one may assign flavor-diagonal couplings to the ALP before EWSB, flavor-changing neutral current interactions will be generated by rotating to the mass basis. In the up- and down-aligned bases, these will be generated in the down- and up-quark sectors, respectively.

We can rewrite the SM fermion currents that the ALP couples to in terms of its vector and axial-vector components as
\begin{align}
\begin{split}
j_{\mathrm{ferm},\mu}
= &\,
\bar{f}_{i,L} \,\big[ k_F  \big]_{ij}\, \gamma_\mu \,f_{j,L} 
+ \bar{f}_{i,R} \,\big[ k_f\big]_{ij} \,\gamma_\mu \,f_{j,R}
=
  \big[ k_F  \big]_{ij} \,\bar{f}_i\, \gamma_\mu \frac{1}{2} \left(1 - \gamma_5 \,\right) f_j 
+ \big[ k_f\big]_{ij} \, \bar{f}_i  \frac{1}{2} \left(1 + \gamma_5 \right) \gamma_\mu \,f_j
\\
\equiv &
  \frac{1}{2} \bar{c}_{f_i f_j} \, \bar{f}_i\, \gamma_\mu \, f_j 
+ \frac{1}{2} c_{f_i f_j} \, \bar{f}_i \,\gamma_\mu \gamma_5\, f_j \, ,
\end{split}
\end{align}
where we have defined
\begin{align}
c_{f_i f_j}  &\equiv \big[ k_f  \big]_{ij} - \big[ k_F  \big]_{ij}\,, & 
\bar{c}_{f_i f_j}  &\equiv \big[ k_f  \big]_{ij} + \big[ k_F \big]_{ij}
 \, .
\end{align}
The flavor-conserving vector currents are conserved in processes mediated by the strong or electromagnetic interactions, making the couplings $\bar{c}_{ff}$ observable only for flavor-changing neutral current processes mediated by the weak interactions~\cite{Bauer:2021wjo}.
For the flavor-diagonal ($i=j$) ALP couplings, we can therefore write~\cref{eq:ferm_lag_EWSB} as
\begin{align}
\mathcal{L}_\mathrm{ferm}^\mathrm{diag}
&= \frac{\partial^\mu a}{2f} \sum_{f=u,d,l} c_{ff} \bar{f} \,\gamma_\mu \gamma_5 f \, ,
\label{eq:ALP_ferm_diag_axial_vector}
\end{align}
where $c_{ff}$ (without indices on the fermions) now refers to the flavor-conserving coefficients only. 

Using integration by parts and equations of motion after EWSB, we can move from the derivative description of~\cref{eq:ferm_lag_EWSB} to the mass-basis equivalent of the chirality-flipping basis of~\cref{eq:ALP_lag_alt}.
\begin{align}
\label{eq:Lferm_eom}
    \mathcal{L}_{\mathrm{ferm}}\simeq
    - i\frac{a}{2f}\left(1+\frac{h}{v}\right)
    \sum_{f=u,d,l}\left(
c_{f_if_j}(m_f^i+m_f^j)\,\bar{f}_i\,\gamma^5f_j
+\bar{c}_{f_if_j}(m_f^i-m_f^j)\,\bar{f}_i\,f_j
    \right) \, ,
\end{align}
where $i,j$ run over the fermion generations. 
The flavor-diagonal couplings take a particularly simple form
\begin{align}
\label{eq:Lferm_diag}
\mathcal{L}_{\mathrm{ferm}}^{\mathrm{diag.}}\simeq
    -i\frac{a}{f}\left(1+\frac{h}{v}\right)
    \sum_{f=u,d,l}
c_{ff}m_f^i\,\bar{f}\gamma^5\,f \,,
\end{align}
corresponding to that of the Yukawa-like operators in~\cref{eq:ALP_lag_alt}.\footnote{The same result can be obtained by performing the appropriate field redefinition discussed in~\cref{sec:interactions} to transform the derivative basis into the chirality-flipping basis and subsequently breaking EW symmetry to obtain the mass-basis interactions.}
The approximate equality here emphasizes the fact that this relation neglects the shifts of ALP-gauge--boson couplings that are induced at one-loop order when moving between the two bases. They can be derived knowing the anomalous contributions to the divergence of the axial currents.
This nevertheless implies that the derivative ALP couplings are classically equivalent to~\cref{eq:Lferm_eom} and therefore lead to the same \emph{tree-level} amplitudes. All such amplitudes mediated by ALP-fermion couplings are therefore proportional to the fermion mass. 

As with all unknown coupling parameters of theories beyond the SM, the ALP-EFT coefficients are subject to theoretical constraints from partial wave unitarity requirements of scattering amplitudes. These have been calculated in Ref.~\cite{Brivio:2021fog} assuming a flavor universal ALP.

\subsection{Chiral Lagrangian}
\label{sec:ALP-chiral}
At energies of the order of $2$~GeV, the only active degrees of freedom are the light quarks $u$, $d$ and $s$, which are confined to hadrons. We should therefore describe the ALP interactions by their coupling to the light mesons $(\pi, K, \eta)$ at and below this energy scale. 
We do not discuss the chiral effective Lagrangian here and refer to~\cite{Bauer:2020jbp} and references therein instead. 

\subsection{Renormalization group effects}
\label{sec:ALP-RGE}
\begin{figure}
    \centering
    \includegraphics[width=0.3\linewidth]{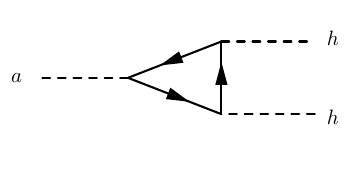}
    \qquad
    \includegraphics[width=0.3\linewidth]{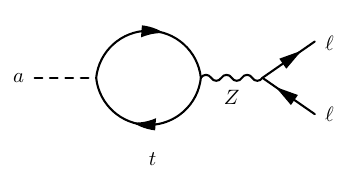}
    \caption{One-loop contributions to the RG running and matching of the ALP-lepton coupling from the ALP-top coupling.}
    \label{fig:RG}
\end{figure}
The couplings of ALPs are probed by various experiments across a wide range of energy scales. 
Therefore, a consistent analysis of the corresponding bounds requires the use of the renormalization group (RG) equations and matching. 
The effective Lagrangian in~\cref{eq:ALP_lag} is assumed to arise from integrating out new heavy particles at a scale $\Lambda$ far above the weak scale. Since we are interested in making predictions at scales around $m_a$, which is assumed to be well below $\Lambda$, we can evolve the Wilson coefficients and operators in the effective
Lagrangian down to the scales of interest by solving their RG equations.

The form of the Lagrangian chosen in~\cref{eq:ALP_lag} (with the couplings $\alpha_i$ with $i=s,L,Y$ pulled out from the definition of the Wilson coefficients) ensures that the ALP couplings to the SM gauge fields do not receive any running contributions~\cite{Chetyrkin:1998mw}
\begin{align}
\beta_{\cGG} & = \frac{\text{d}}{\text{d} \log \mu} \cGG = 0 \,,
&
\beta_{\cWW} & = \frac{\text{d}}{\text{d} \log \mu} \cWW = 0 \,,
&
\beta_{\cBB} & = \frac{\text{d}}{\text{d} \log \mu} \cBB = 0  \, .
\end{align}
However, the ALP-fermion couplings, $\cF$, are scale dependent and satisfy a set of RG equations. 
These have been calculated including two-loop effects in the gauge couplings and at one loop in the Yukawa interactions~\cite{Choi:2017gpf,MartinCamalich:2020dfe,Chala:2020wvs,Bauer:2020jbp,DasBakshi:2023lca,Galda:2025tqs}.
Note that even though certain operators like $\mathcal{O}_H$ can be removed from the basis, they are still required as counterterms when calculating the RG evolution. Any dependence on the shifts $\mathbf{x}_{\sss F}$ necessary to remove $\mathcal{O}_H$ from the basis will drop out in the prediction of physical quantities.

Depending on the energy scale of an observable, we may need to run from the high scale $\Lambda$ to the EW scale, match to the low-energy effective theory, run down to $\sim 2$~GeV and match onto the chiral effective Lagrangian. 
Numerically, it is typically the running from $\Lambda$ to the EW scale which contributes dominantly. 

An interesting feature of the RG evolution is that all fermion couplings, including the lepton couplings $c_{\ell \ell}$, receive large running contributions from the top-quark coupling $c_{tt}$. Therefore, even models without lepton couplings at the high scale can have sizable ALP-lepton couplings at a lower, experimentally accessible, scale. 
The contributions from the running down to the electroweak scale (approximated at one-loop order) and the one-loop matching corrections generated when integrating out the top quark can be written as
\begin{align}
c_{\ell\ell}(\mu) &= c_{\ell \ell}(\Lambda) + 3 \frac{\alpha_t (\mu)}{\alpha_s(\mu)} \left[ 1- \left( \frac{\alpha_s(\Lambda)}{\alpha_s(\mu)} \right)^{1/7} \right] \, c_{tt}(\Lambda) + \, \cdots \, , \\
\Delta c_{\ell \ell}(\mu_\text{EW}) &= -  3 T_3^f \frac{\alpha_t (\mu_\text{EW})}{2 \pi} \log \left( \frac{\mu_\text{EW}^2}{m_t^2} \right) \, c_{tt}(\mu_\text{EW}) + \, \cdots \, ,
\end{align} 
respectively, where the ellipsis refers to contributions from Wilson coefficients other than $c_{tt}$, and $\alpha_t = y_t^2/(4\pi)$ with $y_t\sim1$ denoting the SM top Yukawa coupling. 
Numerically, assuming $c_{\ell\ell}(\Lambda)=0$,
\begin{align}
    c_{\ell\ell}(\mu_\text{EW}) + \Delta c_{\ell \ell}(\mu_\text{EW}) &= (0.130 - 0.019 ) \,  c_{tt}(\Lambda) = 0.111 \, c_{tt}(\Lambda) \, ,
\end{align}
such that the ALP coupling to leptons at the electroweak scale is only an order of magnitude smaller than the ALP-top coupling at the high scale, $\Lambda$.\footnote{This one-loop result is a good approximation of the two-loop result of $0.116 \, c_{tt}(\Lambda)$ in~\cite{Bauer:2020jbp}.}  
We show the diagrams corresponding to the running and matching contributions in~\cref{fig:RG}. The diagram on the left is UV-divergent and requires the operator $\mathcal{O}_H$ as a counterterm. Mapping $\mathcal{O}_H$ back to operators describing ALP-fermion interactions gives rise to the mixing of the top and lepton couplings. This effect has important phenomenological consequences for light ALPs at or below the keV scale, where the coupling to electrons is tightly constrained by astrophysical observables and can imply strong, indirect constraints on other ALP couplings such as $c_{tt}$ via RG running~\cite{Chala:2020wvs}.

\subsection{Effective couplings}
\label{sec:eff-coupl}
Beyond tree level, we can define effective couplings that absorb potential loop-contributions, 
\begin{align}
\label{eq:Ceff}
    C_x^\mathrm{eff} = c_x + (\text{loop\,-\,contributions}),
\end{align} 
where the $c_x$ correspond to the tree-level couplings to fermions and gauge bosons, $c_{ff}$ and $c_{\scriptscriptstyle VV'}$,  as defined in Eqs.~\eqref{eq:ALP_ferm_diag_axial_vector} and \eqref{eq:coup_gauge}, respectively. Note that, like their associated ALP-EFT Wilson coefficients, the effective couplings to gauge bosons retain $\alpha_i/4\pi$ loop normalizations, where $\alpha_i = \alpha, \alpha_s$.
These effective couplings are defined by the loop-level form-factors for the associated three-point amplitudes evaluated on-shell. As such, they cannot be directly used in Feynman diagrams for higher-point amplitudes, where a complete set of loop-diagrams involving the tree-level vertices should be considered.
For a detailed study of one-loop contributions to the effective couplings, we refer to~\cite{Bonilla:2021ufe}. 
 
Here, we only consider the effective ALP-gluon coupling as an example. 
Just like the gluon couplings of the SM Higgs boson, this coupling receives sizable contributions from diagrams with a quark loop 
\begin{equation}
    \label{eq:CGGeff_derivative}
    \Ceffgg
    = \cGG + \frac{1}{2} \sum_q  c_{qq}(m_a) \,B_1(\tau_q)  \, ,
\end{equation}
where $\tau_q = m_a^2/(4 m_q^2)$ and the loop functions are given by\footnote{Note that in the literature there exist different definitions of $\tau$ (which is sometimes defined of the inverse of its definition here) and the function $f$ (which is sometimes defined as the square root of its definition here). }
\begin{equation}
\begin{array}{l}
    B_1(\tau) = 1- \tau^{-1} f(\tau)\,\,, 
\end{array}
\qquad
f(\tau) = 
    \begin{cases}
    \arcsin^2 \sqrt{\tau},   & \tau \leq 1 \, , \\
- \frac{1}{4} \left( \ln \frac{1+ \sqrt{1-1/\tau}}{1- \sqrt{1-1/\tau}} - i \pi \right)^2 , &  \tau > 1 \, .
\end{cases}
\label{eq:B1_loop_func}
\end{equation} 
Evaluating the loop function $B_1$ in the limits of an ALP mass much larger or smaller than a given quark $q$, we find that $B_1(\tau) \approx 1$ for $m_q \ll m_a$ and $B_1(\tau) \approx - m_a^2/(12 m_q^2)$ for $m_q \gg m_a$. 
We can conclude that (very much unlike the SM Higgs coupling) all fermions lighter than $m_a$ contribute significantly to the effective coupling of the ALP to gluons. 

We can alternatively write the effective coupling in terms of the parameters of the Lagrangian in~\cref{eq:ALP_lag_alt} 
\begin{equation}
   \Ceffgg
    = \tcGG +\frac{1}{2}  \sum_q c_{qq} \left[ B_1(\tau_q) -1 \right]  \, ,
\end{equation}
where the $-1$ in the bracket is a result of the different Feynman rules of the fermion couplings. 
Comparing this result with~\cref{eq:CGGeff_derivative} and using the relation between $\cGG$ and $\tcGG$ from~\cref{eq:relation_derivative_pseudoscalar} in the mass basis
\begin{align}
\label{eq:tcgg_correction}
    \tcGG &= \cGG + \frac{1}{2} \text{Tr}\left[ \cu + \cd - 2 \cQ \right] = \cGG + \frac{1}{2} \sum_q c_{qq} \, ,
\end{align}
we can see that the effective coupling is the same in both bases. 
We note, again, that the two bases are equivalent up to one-loop effects in the fermion couplings. 

This discussion highlights an important point of distinction between the fermionic couplings of a shift-symmetric ALP from those of a generic pseudoscalar particle. Had we started from a non-zero fermion coupling in the chirality-flipping basis, we would have observed the non-decoupling behavior, $\Ceffgg\to\mathrm{constant}$, when taking $m_q\to\infty$, analogous to the Higgs boson case. Taking, instead, a shift-symmetric (derivative) fermion coupling for the ALP predicts a decoupling behavior, $\Ceffgg\to0$, in this limit. The loop-level correction to $\tcGG$ in~\cref{eq:tcgg_correction} compensates the non-decoupling of the loop-function of the chirality-flipping coupling to recover the same prediction in the shift-symmetric limit. Similarly, shift-symmetric fermion couplings yield corrections to ALP-photon couplings that decouple in the limit of large $m_q$ or $m_a\to 0$. Fermion coupling contributions to the effective couplings to massive EW gauge bosons do not decouple in general and the limit $m_a\to0$ is of course not possible with all particles on-shell~\cite{Bonilla:2021ufe}.

\section{Production at colliders}\label{sec:production}
In this section, we will discuss various on-shell production modes for ALPs at colliders. As we will see, ALPs can be produced in many of the same ways as the SM Higgs boson. However, the fact that ALP interactions are described by an EFT with higher derivative interactions leads to ALP production modes displaying a characteristic energy growth compared to a general spin-zero particle. This means that we can exploit the high center-of-mass energies of collider experiments to efficiently produce and search for ALPs. In this section, we will focus on single ALP production processes, although we note that ALP pair production is also a potentially interesting avenue for investigation. This process would inherently arise at dimension-six in the ALP-EFT and we will not discuss it further in this review (see~\cite{Alves:2021puo,Bauer:2022rwf,Cheung:2024lhn} for recent studies). Different production modes will be sensitive to different ALP couplings with SM fields and we will proceed by discussing several possible production channels that could be used to probe different sectors of the ALP interactions introduced in~\cref{sec:interactions}. 

\subsection{Gauge boson couplings}
\label{subsec:prod_gauge}
The fact that the ALP is a SM singlet pseudoscalar dictates a characteristic structure for its coupling with gauge bosons:
\begin{align}
\label{eq:L_aXXp}
    \mathcal{L}_{aXX^\prime} = \gXXp \frac{a}{f}X^{\mu\nu}\widetilde{X}^\prime_{\mu\nu};\quad X_{\mu\nu},X^\prime_{\mu\nu}\in\left\{F_{\mu\nu},Z_{\mu\nu},W^\pm_{\mu\nu},G^A_{\mu\nu}\right\},
\end{align}
which leads to the following Feynman rule for the three-point $a$-$X$-$X^\prime$ vertex, with all momenta  incoming:
\begin{align}
\label{eq:gaXXp_FR}
\includegraphics[width=0.22\textwidth]{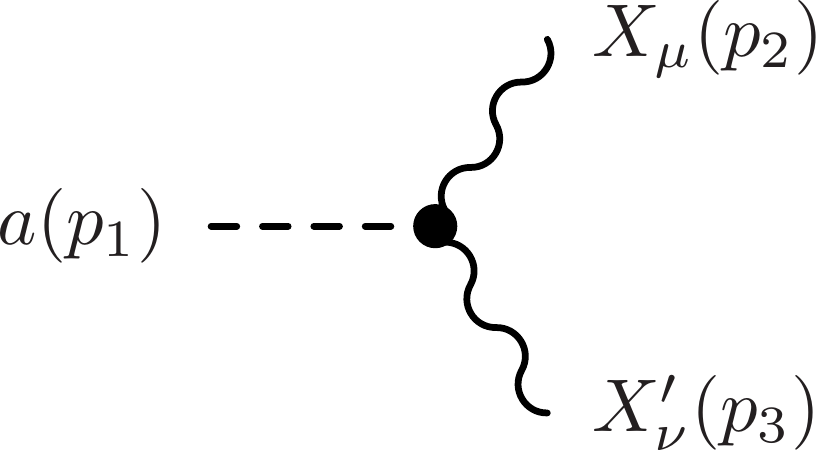}\quad\raisebox{0.95cm}{$:\qquad   2 \,i\,(1+\delta_{\sss XX^\prime})\dfrac{\gXXp}{f} \epsilon_{\mu\nu\rho\sigma} \,p_2^\rho\, p_3^\sigma\,,$}
\end{align}
where $\epsilon_{\mu\nu\rho\sigma}$ is the totally antisymmetric Levi-Civita tensor and the Kronecker delta, $\delta_{\sss XX^\prime}$, accounts for the extra symmetry factor when $X$ and $X^\prime$ are identical. The vertex depends on the external gauge boson momenta, $p_2$ and $p_3$, and $\epsilon_{\mu\nu\rho\sigma}$ appears because of the \CP-odd nature of the ALP. This interaction can mediate a number of ALP production modes in high-energy particle collisions.

\begin{center}
\begin{figure}[h!]
    \centering
    \includegraphics[width=0.22\linewidth]{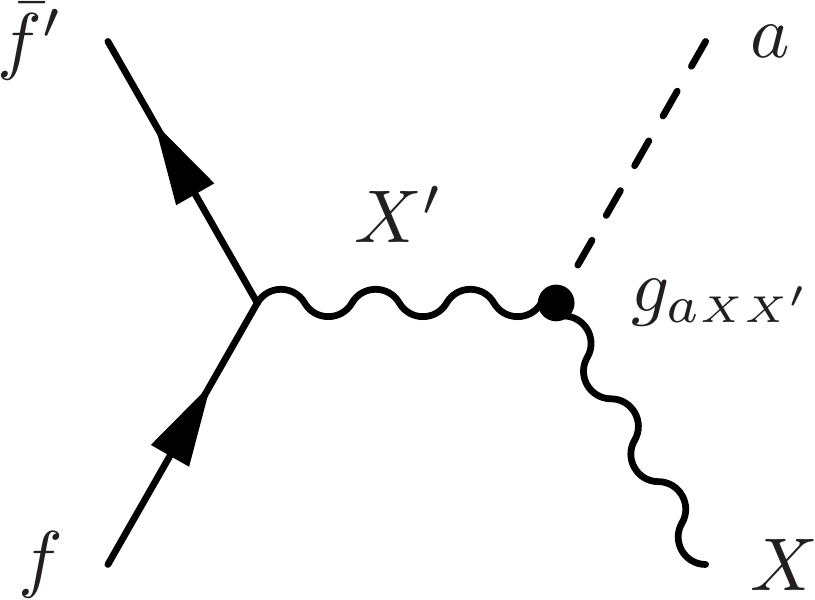}
    \caption{
    \label{fig:associated_production_X}
    Feynman diagram for fermion-anti--fermion initiated ALP production in association with a SM gauge boson, $X$, mediated by the ALP gauge-boson coupling in~\cref{eq:L_aXXp,eq:gaXXp_FR}.}
\end{figure}    
\end{center}

Perhaps the most widely-studied production mode is that ALP production in association with a gauge boson initiated by fermion-anti--fermion scattering, $f\bar{f}^\prime\to a + X$, as depicted in Feynman diagram of~\cref{fig:associated_production_X}.
For a fixed center-of-mass energy, this process predicts a mono-energetic ALP recoiling against a gauge boson in the final state. As an example, an ALP with a photon interaction of the type shown in~\cref{eq:gaXXp_FR} would give rise to an ALP$+\gamma$ final state through the annihilation of any electrically charged fermion-anti--fermion pairs.
The squared matrix element for this scattering process in terms of the mass-basis coupling of~\cref{eq:LEW_mass} is
\begin{align}
    \label{eq:ME_ffxA}
    \left|\mathcal{M}\left(f(p_1)\bar{f}(p_2)\to a(p_3)\gamma(p_4)\right)\right|^2 = 16 \,\gaa^2 \,e^2 Q_f^2\,\frac{s}{f^2}\left(\frac{t^2+u^2}{s^2}\right),
\end{align}
where $e$ and $Q_f$ are the electromagnetic coupling and the electric charge of the fermion, respectively, and $s,t,u$ are the usual Mandelstam variables defined in terms of the external momenta, $p_i$. The amplitude squared displays a characteristic energy growth $\sim s/f^2$, while the quantity in brackets contains the dependence on the scattering angle between the ALP and the initial state fermion. The growth with the square of the energy gives a total, spin-averaged cross section that tends to a constant in the high-energy limit,
\begin{align}
\label{eq:xs_ffxA}
\sigma_{f\bar{f}\to a+\gamma}(s) = 
  \frac{\,e^2 Q_f^2}{6\pi}\frac{g^2_{a\gamma\gamma} }{f^2}\left(1-\frac{m_a^2}{s}\right)^3\,.
\end{align}
This behavior can be inferred on dimensional grounds, since $2\to2$ matrix elements are dimensionless and we have a factor of $1/f$ in the interaction vertex.

The unique interactions of ALPs predict potentially large rates for production in association with a photon, that are relatively enhanced at high energies compared to typical SM backgrounds, which tend to fall steeply with the center-of-mass energy. 
High-energy colliders are therefore a promising place to search for ALPs that couple with the photon, where we would expect signatures involving boosted ALPs recoiling against an energetic photon in the final state, as shown in~\cite{Masso:1995tw,Kleban:2005rj,Mimasu:2014nea}.
Many of the collider bounds from the LEP, Belle II, Babar and BES  III experiments shown in~\cref{fig:AxionLimits} rely on this production process.
The same conclusions can be drawn for ALP production in association with other SM gauge bosons. 
As discussed in~\cref{sec:interactions}, the three couplings $\tilde{c}_{BB}$, $\tilde{c}_{WW}$ and $\tilde{c}_{GG}$ control the ALP-gauge--boson interactions and can mediate production in association with a photon, $Z$-boson, $W$-boson or gluon, with the same high-energy dependence~\cite{Mimasu:2014nea,Brivio:2017ije,Haghighat:2020nuh,Wang:2021uyb,Ren:2021prq,Ghebretinsaea:2022djg,Cheung:2024qge,Aiko:2024xiv}. 
For example, LHC experiments have performed searches for high-mass ALPs in triboson final states mediated by these production processes and the subsequent decay of the ALP into a pair of EW gauge bosons~\cite{CMS:2019mpq,CMS:2025oey}, as well as searches for boosted di-jet resonances that could probe the quark and/or gauge boson couplings~\cite{ATLAS:2018hbc,CMS:2019xai,CMS:2019mcu,ATLAS:2024bms,ATLAS:2024qqm}. In the case of photon- or $Z$-associated production, the process can be mediated in the $s$-channel by either a Z boson or a photon. Therefore, for ALP-$\gamma$ production, there is a potentially resonant channel when $m_a<m_Z$, which can be described by on-shell $Z$-boson production decaying via the $a\gamma$ channel, as we discuss later in~\cref{sec:decay}. 

The other main channel by which the ALP can be produced directly from its coupling to gauge fields is via the fusion of two vector bosons. In this case, the on-shell ALP production amplitude-squared is
\begin{align}
\label{eq:XXp_to_a}
\left|\mathcal{M}\left(X(p_1)X^\prime(p_2)\to a(p_3)\right)\right|^2 &=
2(1+\delta_{XX^\prime})^2\frac{\gXXp^2}{f^2}
m_a^4\,\lambda\left(\frac{m_X^2}{m_a^2},\frac{m_{X^\prime}^2}{m_a^2}\right),
\end{align}
with $\lambda(z, r) = (1 -z -r )^2 - 4 z r $.\footnote{One could also replace $\gXXp$ with $C_{\scriptscriptstyle XX^\prime}^\mathrm{eff}$ in~\cref{eq:XXp_to_a}}
This amplitude is applicable to production via gluon-gluon fusion (ggF) at hadron colliders, mediated by $\cGG$; we refer to, \emph{e.g.},  Refs.~\cite{Jaeckel:2012yz,Mariotti:2017vtv,CidVidal:2018blh,ATLAS:2022abz,ATLAS:2024bjr} for phenomenological and experimental studies in the diphoton decay channel, or Ref.~\cite{Anuar:2024qsz} for a study of heavy ALPs decaying to $t\bar{t}$. 
In general, ALPs produced via gluon fusion can be probed by a wide array of resonance searches performed at the LHC thus far, depending on their preferred decay mode.

\begin{figure}[h!]
  \centering
    \begin{subfigure}{0.24\textwidth}
        \raisebox{0.3cm}{\includegraphics[width=\textwidth]{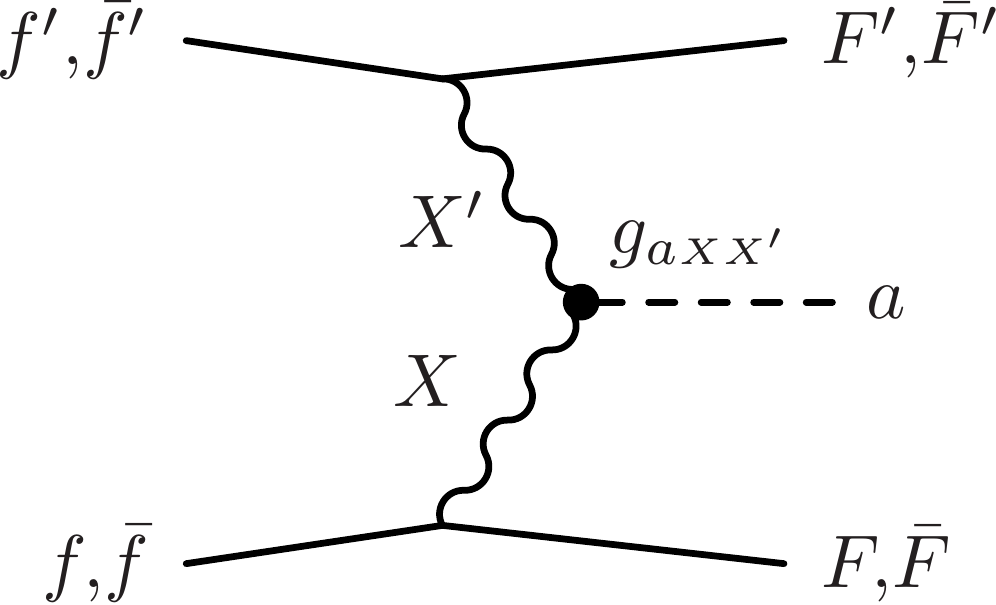}}
          \caption{\label{fig:VBF_ff}}
    \end{subfigure}
    \hspace{2cm}
    \begin{subfigure}{0.24\textwidth}
        \includegraphics[width=\textwidth]{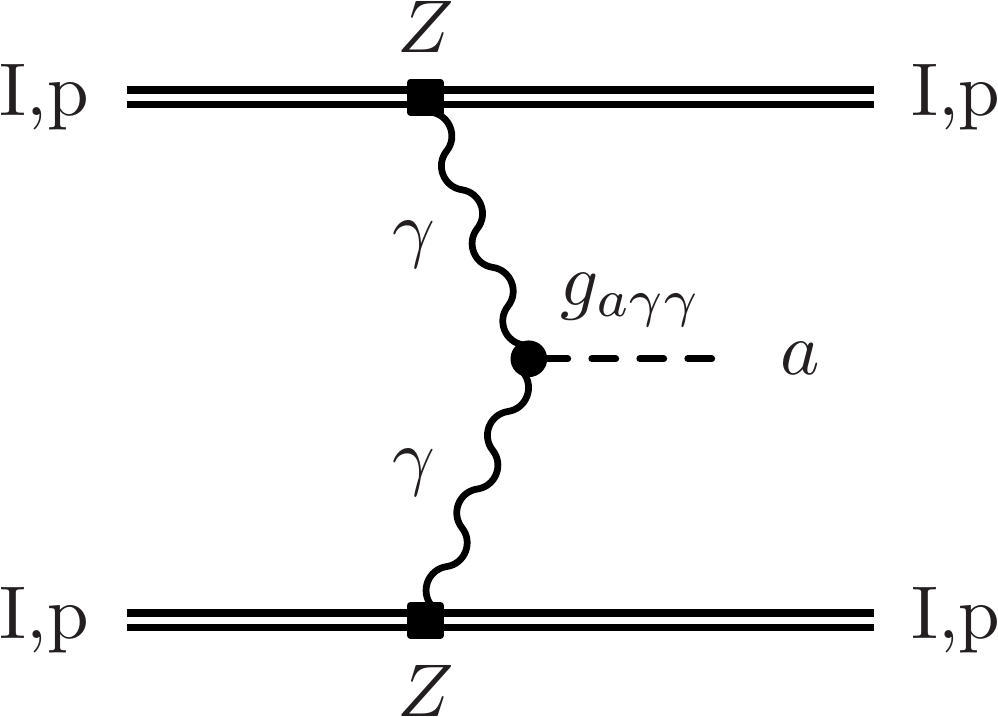}
        \caption{\label{fig:VBF_UP}}
    \end{subfigure}
\caption{
\label{fig:VBF}
Feynman diagrams depicting ALP production via VBF. 
(a) Partonic contribution to non-exclusive VBF production. The emitting legs can be any fermion ($f^{(\prime)},F^{(\prime)}$) or anti-fermion ($\bar{f}^{(\prime)},\bar{F}^{(\prime)}$) and the intermediate vector bosons, $X^{(\prime)}$, can be any force-carrier of the SM ($\gamma/Z/W^\pm/g$).  
(b) Exclusive ALP production via photon fusion in the ultra-peripheral collisions of heavy ions ($I$) or (anti-)protons ($p$) of electric charge, $Z$. The scatterers remain intact in this process.
}
\end{figure}

The other gauge couplings can also mediate on-shell ALP production if the gauge bosons are emitted quasi-collinearly from the colliding particles, as is the case for Higgs boson production via vector boson fusion (VBF)~\cite{Yue:2021iiu,Florez:2021zoo}.~\Cref{fig:VBF_ff} depicts the Feynman diagram for this process in partonic scattering.
Here the gauge bosons are not precisely on-shell, but the rate can be estimated using the equivalent photon approximation~\cite{Fermi:1924tc,vonWeizsacker:1934nji,Landau:1934zj} and its generalization to massive gauge bosons~\cite{Dawson:1984gx}, or the full $2\to3$ process can be directly simulated via Monte Carlo tools. 
This type of production process is possible for lepton, proton-(anti)proton and heavy ion collisions or combinations thereof. In contrast with VBF Higgs production in the SM, there can  also be a sizable photon contribution. In the case of lepton colliders, the neutral and charged VBF channels are easily distinguished as the former retains the initial-state leptons in the final state while the latter yields neutrinos, leading to an ALP plus missing-energy signature.

Ultra-peripheral heavy ion collisions have been shown to be particularly sensitive to the ALP-photon coupling, as the photon luminosity receives a coherent enhancement due to the large electric charge of the ions, $Z$, with the production rate scaling like $Z^4$~\cite{Knapen:2016moh}.~\Cref{fig:VBF_UP} depicts the Feynman diagram for this type of process, where the initial state scatterers remain intact.  
These searches have been performed at the LHC~\cite{ATLAS:2017fur,CMS:2018erd,ATLAS:2019azn,ATLAS:2020hii,CMS:2024bnt}, targeting ALP decays into a pair of photons.
This process was recently used to observe light-by-light scattering ($\gamma\gamma\to\gamma\gamma$) in the SM, which constitutes the main background for the ALP search. Beyond $m_a\sim 100$ GeV, the coherent enhancement of the photon flux is lost as the photons begin to resolve the nuclear structure of the ion. 
The analogous process for proton-proton collisions can extend the sensitivity to higher masses, since the length scale at which the proton structure is resolved is much shorter~\cite{Baldenegro:2018hng,dEnterria:2021ljz}. In order to access the ultra-peripheral regime, the protons are tagged in the final state using dedicated spectrometers placed a few hundred meters downstream of the collision point. 
Although the photon luminosity itself is no longer enhanced by a large electric charge, the large available luminosity delivered by the LHC experiment compensates to a degree~\cite{ATLAS:2023zfc,TOTEM:2023ewz}.

\begin{figure}[h!]
  \centering
    \begin{subfigure}{0.24\textwidth}
        \includegraphics[width=\textwidth]{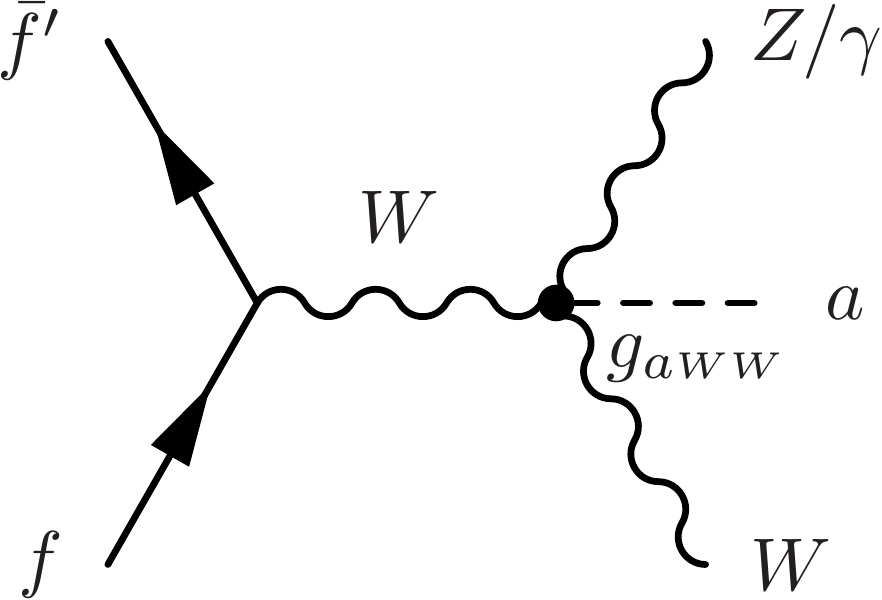}
          \caption{\label{fig:ffp_axWV}}
    \end{subfigure}
    \hspace{2cm}
    \begin{subfigure}{0.24\textwidth}
        \includegraphics[width=\textwidth]{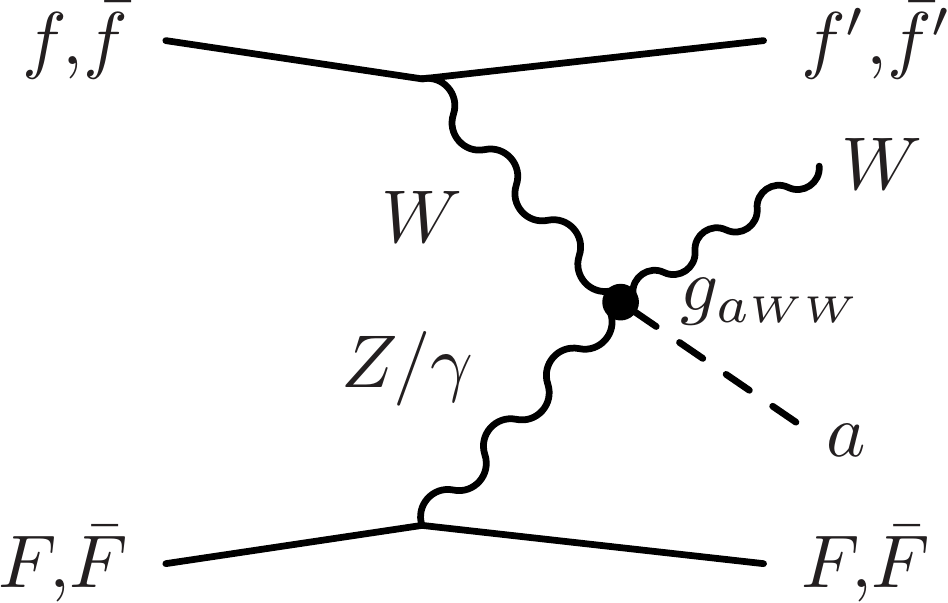}
        \caption{\label{fig:VBF_axW}}
    \end{subfigure}
\caption{
\label{fig:TGC}
Feynman diagrams for the rare ALP-production modes sensitive to the four-point interactions with gauge bosons that only depend on $\gWW$. 
(a) Production in association with a $W$- and a $Z$-boson or a photon. 
(b) Production via VBF in association with a $W$-boson.
}
\end{figure}

The ALP couplings to EW gauge bosons can mediate other direct production processes, in association with a number of other final states. Some notable examples include production in association with a gauge boson pair, $f\bar{f}^\prime \to a +X X^\prime$, and vector boson fusion in association with a gauge boson, $f\bar{f}^\prime \to a+F\bar{F}^\prime X$. In the case where $W$ bosons appear in intermediate or final states, both processes are sensitive to the four-point interactions arising from the non-Abelian nature of $S\!U(2)$, shown in the middle line of~\cref{eq:LEW_mass}.~\Cref{fig:TGC} shows two example Feynman diagrams for these processes. At dimension five, these interactions are not independent of the lower point couplings but could be decoupled from $\gWW$ at higher orders in the ALP EFT, starting at dimension seven.
\subsection{Fermion couplings}
\label{subsec:prod_fermion}
As discussed in~\cref{sec:interactions}, ALP-fermion couplings are proportional to the fermion mass.
This makes ALP-fermion couplings more challenging to probe experimentally, particularly those of the lighter fermions. 
In fact, all of the processes mediated by the gauge boson couplings discussed in~\cref{subsec:prod_gauge} also have contributions in principle from the various $\cF$. However, in practice they involve the emission of an ALP from a light lepton or quark and are therefore highly suppressed,  justifying their omission.

\begin{figure}[h!]
  \centering
    \begin{subfigure}{0.24\textwidth}
        \includegraphics[width=\textwidth]{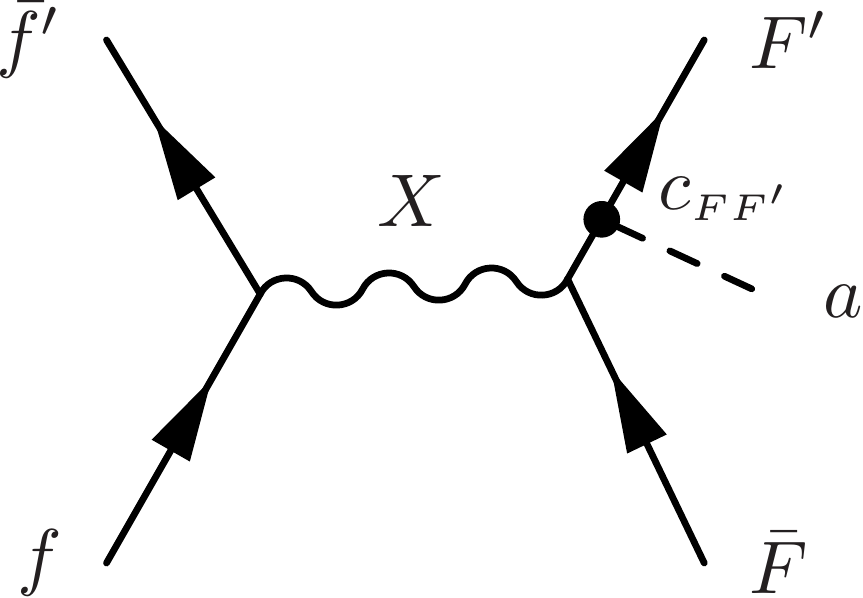}
          \caption{\label{fig:ffp_aFFp}}
    \end{subfigure}
    \hspace{2cm}
    \begin{subfigure}{0.24\textwidth}
        \includegraphics[width=\textwidth]{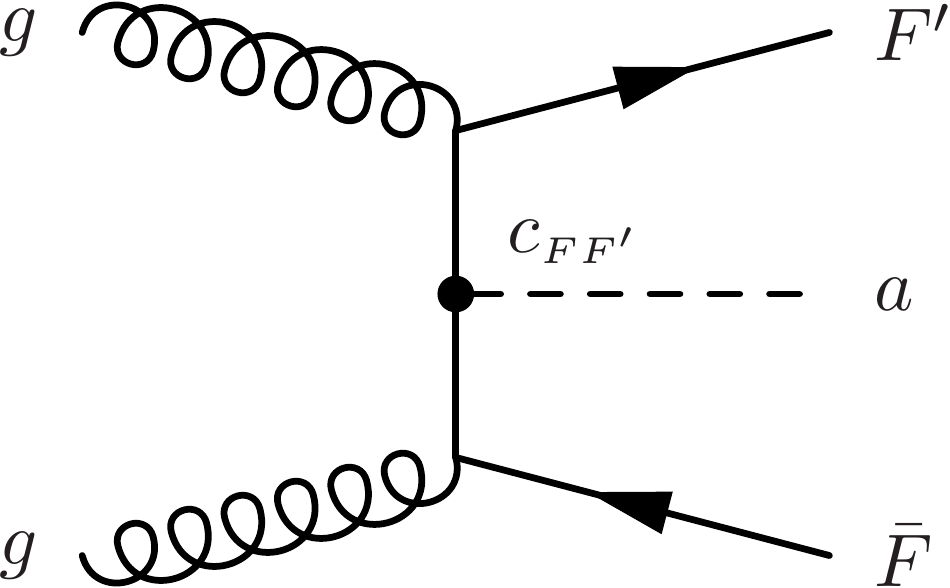}
        \caption{\label{fig:gg_aFFp}}
    \end{subfigure}
\caption{
\label{fig:ffa}
Feynman diagrams for ALP production in association with a fermion-anti--fermion pair via an (a) fermion-anti--fermion initial state and a (b) gluon-gluon initial state relevant for production in association with a quark-anti--quark pair at hadron colliders.
}
\end{figure}

Therefore, like the SM Higgs, the most promising production modes for probing ALP-fermion couplings are those involving the heaviest fermions. The processes that achieve direct sensitivity to the ALP-fermion couplings are ALP production in association with fermions. Most notably we have production in association with a fermion-anti--fermion pair, $f\bar{f}^\prime \to a+F\bar{F}^\prime$ (or $gg \to a+F\bar{F}^\prime$ in the hadron collider case, where $F,F^\prime$ are quarks) as shown by the Feynman diagrams in~\cref{fig:ffa}.
In practice, $t\bar{t}$ and potentially $b\bar{b}$ pairs offer the most promising sensitivity at hadron colliders. Some phenomenological studies and experimental searches for $pp \to a+t\bar{t}$ production have been performed targeting the $ b{\bar{b}}$~\cite{Blasi:2023hvb,ATLAS:2025oxn} and leptonic~\cite{CMS:2024ulc} decay modes of the ALP, as well as for long-lived ALP~\cite{Rygaard:2023dlx,Esser:2023fdo}. A related channel is ALP production in association with, \emph{e.g.}, a single top quark. Considering all of the single-top--quark production channels in the SM, it is possible to emit an ALP from any of the intermediate or external particles. Being an EW-induced process, it will be sensitive not only to the left-handed ALP-top and ALP-bottom couplings, but additionally to the $W$-boson and gluon couplings~\cite{Ebadi:2019gij,Hosseini:2024kuh,Cheung:2024wve,Eberhart:2025lyu}. 
For a collection of collider bounds on ALP-lepton couplings, see~\cite{Bauer:2017ris,Alda:2024cxn}.

Besides the tree-level processes discussed so far, another way to access ALP-fermion couplings at colliders is by their contributions to loop-level processes such as gluon fusion where, as discussed in~\cref{sec:eff-coupl}, ALP-gauge--boson couplings receive one-loop contributions from all fermions. Ref.~\cite{Anuar:2024qsz}, for example, studies the sensitivity of a heavy, top-philic ALP produced in gluon fusion and decaying into a top-antitop pair. In general, the effective ALP-gauge--boson couplings offer sensitivity to derivative couplings to light-fermions, which do not decouple in the same way that Yukawa couplings of the Higgs bosons do. 

\subsection{Production cross sections at colliders}
Here we present some illustrative cross sections for ALP production at colliders. These were computed at tree level and parton level using \texttt{Madgraph5\_aMC@NLO3.5.3} \cite{Alwall:2014hca} and a \texttt{FeynRules}~\cite{Alloul:2013bka} implementation of the ALP-EFT. \texttt{PDF4LHC21} parton distribution functions~\cite{PDF4LHCWorkingGroup:2022cjn} are used, and jets are restricted to have $p_T>20$ GeV and pseudorapidity $|\eta|<5$, where relevant. All rates are shown as a function of $m_a$ between 10 GeV and 1 TeV, with an illustrative coupling value of $c_{x}/f=1\,\mathrm{TeV}^{-1}$, where $c_{x}$ is a coupling parameter that mediates the process at tree-level. 

In all cases, possible contributions from resonant $W$ or $Z$ decays to overlapping final states are removed, to highlight the intended production processes. For example, VBF production at the LHC $(pp\to a+jj)$ could be polluted by contributions from production in association with a hadronically-decaying $W$ or $Z$ boson. The latter would contribute in a completely different region of phase space and could therefore be easily distinguished from VBF production, which is characterized by energetic jets with a large invariant mass and rapidity separation. Analogous overlaps between VBF and associated production modes at lepton colliders are removed. 

\begin{figure}[h!]
\centering
\includegraphics[width=0.5\textwidth]{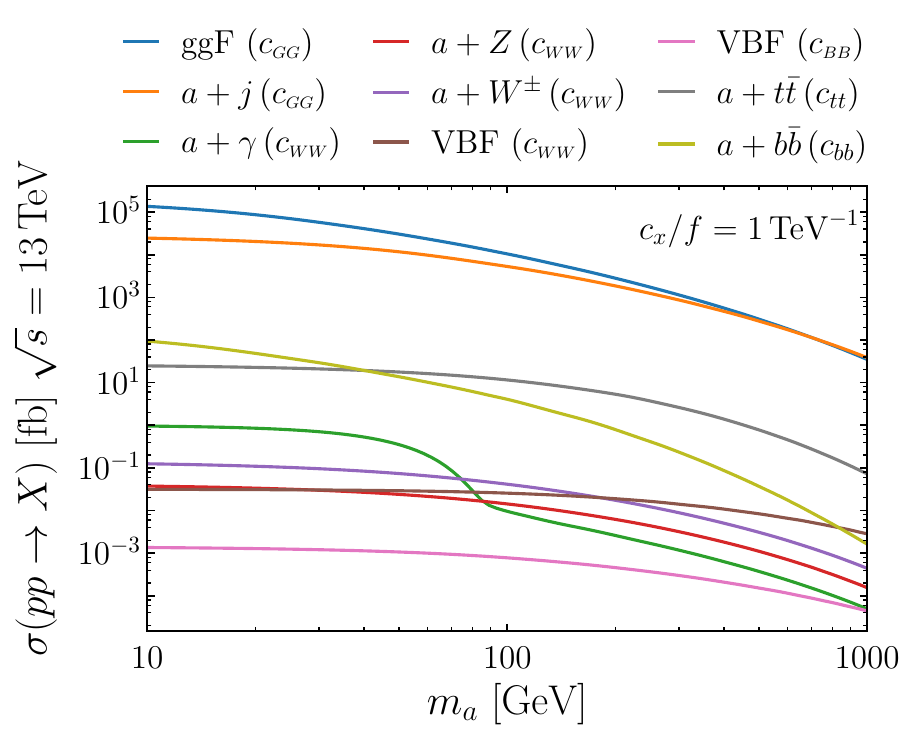}
    \caption{
    \label{fig:prod_pp}
    Cross sections in femtobarn for ALP production modes in 13 TeV proton-proton  collisions. For each process, a single relevant ALP coupling, indicated in brackets, that mediates the process at tree-level has been switched on with a value of $c_x/f=1$ TeV$^{-1}$.
    }
\end{figure}

\Cref{fig:prod_pp} plots cross sections for various production modes at the 13 TeV LHC including ggF, VBF and production in association with a jet, a SM gauge boson or a pair of top or bottom quarks. The cross sections span many orders of magnitude across the different processes, and each one drops by several orders of magnitude across the $m_a$ range.
For couplings of order one, the gluon-initiated processes are the largest, followed by those that depend on the fermion couplings. The processes mediated by the EW gauge boson couplings are the rarest, owing partly to the normalization of the operators with the $\alpha_i/4\pi$ factors in~\cref{eq:ALP_lag}.  VBF rates are shown both for $\cBB$ and $\cWW$ switched on, as the former mediates only neutral weak boson fusion, while the latter also includes $W$-boson fusion, and leads to much larger cross sections. Despite the analogous normalization factor, $\cGG$-mediated processes still dominate over the fermion couplings thanks to the large gluon luminosity at the LHC. The $a+\gamma$ channel drops sharply at $m_a=m_Z$ when it can no longer be mediated by an on-shell $Z\to a\gamma$ decay (see~\cref{sec:prod_in_decays}). VBF falls off more slowly with $m_a$ and becomes the dominant production mode for $\cWW$ at $m_a\sim180$ GeV. Although it is not shown in the figure, VBF also becomes the dominant production mode at $m_a\sim130$ GeV for $\cBB$. On the fermion couplings, it is interesting to note the mass dependences of the $b\bar{b}$ and $t\bar{t}$ associated production processes. Even though the ALP-fermion couplings predict $a+b\bar{b}$ to suffer a relative suppression of $m_b^2/m_t^2\sim10^{-3}$ with respect to  $a+t\bar{t}$, it still wins out at low $m_a$ due to the significantly larger phase space and the larger parton luminosities at lower Bjorken-$x$. 

\begin{figure}[h!]
\centering
\includegraphics[width=0.8\textwidth]{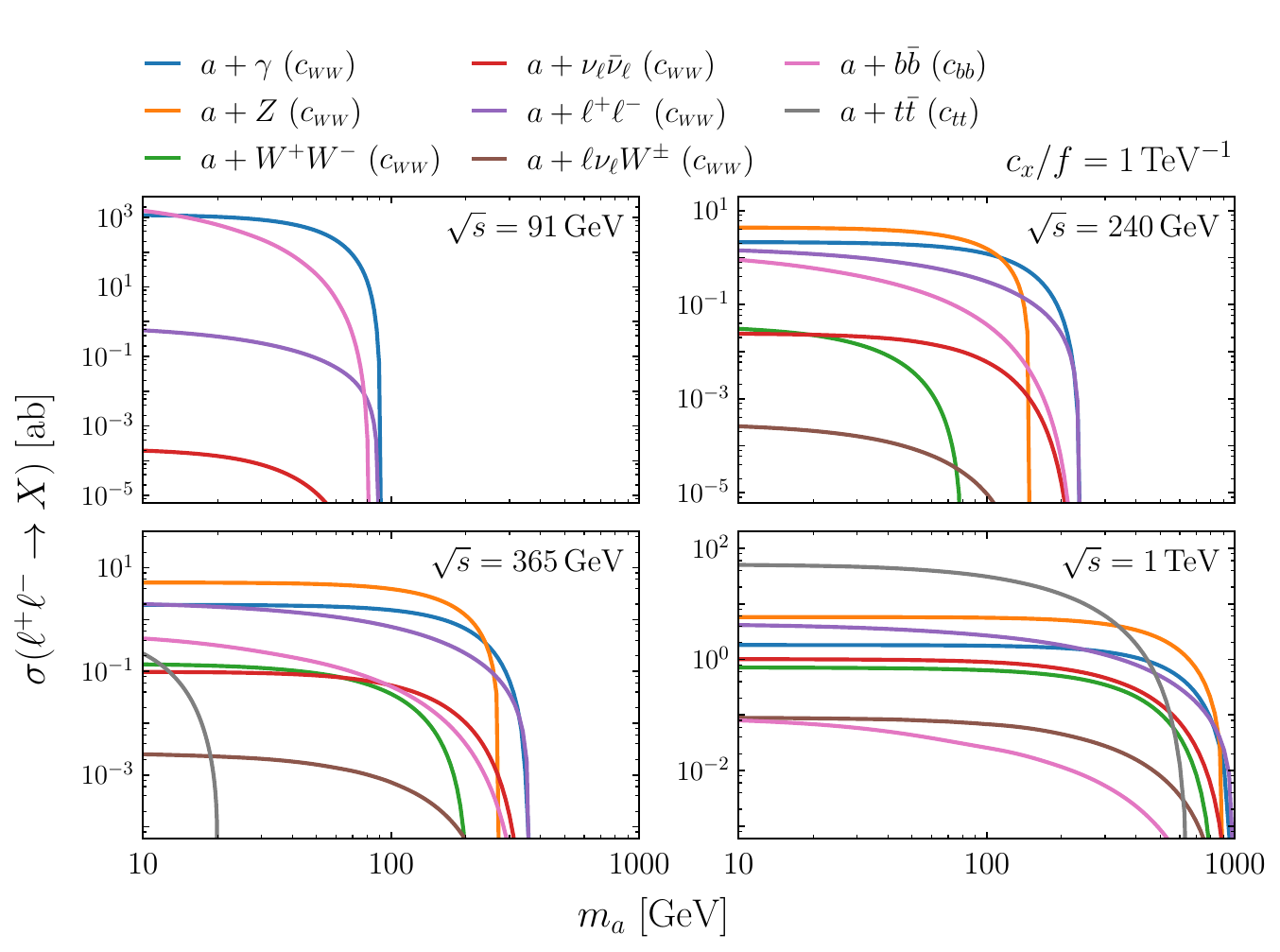}
    \caption{
    \label{fig:prod_ee}
    Cross sections in attobarn for ALP production modes in lepton anti-lepton collisions at various center-of-mass energies. For each process, a single relevant ALP coupling, indicated in brackets, that mediates the process at tree-level has been switched on with a value of $c_x/f=1$ TeV$^{-1}$.
    }
\end{figure}

~\Cref{fig:prod_ee} shows some production modes at lepton colliders with center-of-mass energies corresponding to the $Z$-pole (91.2 GeV), the Higgsstrahlung peak (240 GeV), just above the $t\bar{t}$ threshold (365 GeV) and 1 TeV. The energies are illustrative of possible running points for proposed future $e^+ e^-$ colliders, see~\cite{Bauer:2018uxu,CLIC:2018fvx,Buttazzo:2018qqp,Inan:2020aal,Cacciapaglia:2021agf,Zhang:2021sio,Calibbi:2022izs,Inan:2022rcr,Liu:2022tqn,Tian:2022rsi,Wang:2022ock,Cheung:2023nzg,RebelloTeles:2023uig,Yue:2022ash,Yue:2023mjm,Yue:2024xrc,Bao:2025tqs} for recent ALP studies. These include $Z$ and $\gamma$ associated production, neutral ($a+\ell^+\ell^-$) and charged ($a+\nu_\ell\bar{\nu}_\ell$) VBF, $b\bar{b}$ and $t\bar{t}$ associated production, and -- as examples of processes that are sensitive to the four-point ALP-gauge boson interactions -- $W^+W^-$ associated production and VBF production in association with a $W$-boson (see~\cref{fig:TGC}). While neutral VBF and $\gamma/Z$ associated production depend on both $\cWW$ and $\cBB$, the remaining EW production processes are only sensitive to $\cWW$. To avoid poles in VBF amplitudes mediated by photons, a cut of $p_T>1$ GeV was imposed on final state charged leptons.
 At the $Z$-pole, the dominant channels come from on-shell $Z\to a\gamma$ and $Z\to ab\bar{b}$ decays. The loop-level normalization of $\cWW$ in the former and the three-body phase-space suppression of the latter lead to similar partial widths. Weak boson fusion processes also make up a subdominant component at $\sqrt{s}=$91 GeV. 
 At 240 GeV, production in association with $Z/\gamma$ becomes the most important and the neutral weak boson fusion grows in importance thanks to its photon-fusion component. Although not shown in the figure, $a+\gamma$ production is most important for $\cBB$. $a+b\bar{b}$ rates decrease with center-of-mass energy due to their lack of energy growth with respect to the weak production modes. As the energy increases further, the $a+t\bar{t}$ channel opens and dominates at the highest energies, thanks to the large top Yukawa enhancement. The relative importance of the weak-boson-fusion modes increases with $\sqrt{s}$, and neutral VBF eventually becomes the most important production mode for $\cWW$ and $\cBB$, particularly at lower $m_a$. 
Future muon colliders~\cite{Bao:2022onq,Haghighat:2021djz, Han:2022mzp, Inan:2022rcr,Lu:2023ryd,Chigusa:2023rrz,Calibbi:2024rcm,Batell:2024cdl,Li:2024zhw,Chigusa:2025otr} or electron-ion colliders~\cite{Yue:2019gbh,Oliveira:2021qtq,Liu:2021lan,Davoudiasl:2021mjy,Mosala:2023sse,Balkin:2023gya,Yue:2023mew,Davoudiasl:2024fiz,Giang:2025wif} can also provide relevant constraints on ALPs, but are not further discussed in this review.

\subsection{Production in decays}
\label{sec:prod_in_decays}
As already mentioned above, ALPs can also be produced in the decay of SM particles, for instance in flavor-violating meson decays as well as decays of the Higgs and $Z$ boson. The large statistics collected thus far at flavor experiments, LEP and the LHC allow in many cases for precise measurements of total widths and searches for exotic decay modes involving the ALP.
Limits on exotic branching fractions of the Higgs and $Z$-bosons hence result in important constraints on the couplings that induce these decays.  
The decay widths for the channels $Z \to \gamma a$, $h \to Z a$ and $h \to aa$ can be written in terms of the effective ALP couplings of~\cref{eq:Ceff} as~\cite{Bauer:2017ris}:
\begin{align}
\begin{split}
\Gamma(Z \to \gamma a ) &= \frac{m_z^3}{96 \pi^3 f^2} \frac{\alpha^2}{\sw^2 \cw^2} \big|\CeffaZ\big|^2 \left( 1 - \frac{m_a^2}{m_Z^2} \right)^3 \, , \\
\Gamma(h \to Z a ) &= \frac{m_h^3}{16 \pi f^2}  \big|\CeffZh\big|^2  \lambda^{3/2}\left( \frac{m_Z^2}{m_h^2} , \frac{m_a^2}{m_h^2} \right)  \, ,\\
\Gamma(h \to a a ) &= \frac{m_h^3 v^2}{32 \pi f^4} \big|\Ceffah\big|^2 \left( 1 - \frac{2m_a^2}{m_h^2} \right)^2 \sqrt{1 - \frac{4 m_a^2}{m_h^2}} \, .
\end{split}
\end{align}

While an ALP coupling to the $Z$ boson and a photon exists at tree-level at dimension-five, the effective $h$-$a$-$a$ coupling is only induced at loop level or at mass dimension six via the operator corresponding to $\cHH$. 
The decay $h \to Z a$ first arises at dimension seven or via top-loop contributions. Note that the redundant operator $\mathcal{O}_H$ does not contribute to a $h$-$Z$-$a$ coupling as discussed in Section~\ref{sec:interactions}. 
We refer the reader to~\cite{Bauer:2020jbp} for an explicit representation of the loop contributions to the effective couplings. 
ATLAS and CMS provide bounds on $h\to aa$ ~\cite{CMS:2018qvj,CMS:2018zvv,CMS:2018nsh,CMS:2019spf,CMS:2020ffa,ATLAS:2021hbr, ATLAS:2023ian,CMS:2024uru,ATLAS:2024vpj}
and $h \to Z a$~\cite{ATLAS:2023etl}
coming from direct searches for exotic Higgs decays in various final states such as corresponding to different ALP decay modes. The above effective couplings can also be constrained by the total $Z$-boson-width measured at LEP as well as the unmeasured Higgs branching fraction inferred collectively from LHC signal-strength measurements.

\begin{figure}[t]
\centering
\includegraphics[width=0.28\textwidth]{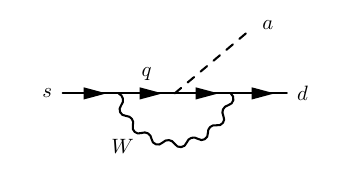}
\quad
\includegraphics[width=0.28\textwidth]{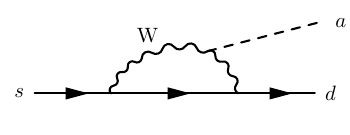}
    \caption{Sample Feynman diagrams contributing to effective flavor-violating ALP couplings.}
    \label{fig:feyn_Ktopi}
\end{figure}
For masses below  $\sim 5 \text{ GeV}$, flavor experiments probing rare or exotic meson decays provide some of the most stringent constraints on ALPs~\cite{Bardeen:1978nq,Antoniadis:1981zw,Georgi:1986df,Krauss:1986bq,Izaguirre:2016dfi,Dolan:2017osp,Dobrich:2018jyi,Merlo:2019anv,Gavela:2019wzg,Ertas:2020xcc,Gori:2020xvq,Ishida:2020oxl,Bertholet:2021hjl,Gallo:2021ame,Bauer:2021wjo,Bauer:2021mvw,Bandyopadhyay:2021wbb,Guerrera:2021yss,Guerrera:2022ykl,DiLuzio:2022tbb,Cornella:2023kjq,DiLuzio:2024jip,Alves:2024dpa,Cavan-Piton:2024pqp,Berger:2024xqk,Alda:2025uwo}. 
These precision tests rely on ALP contributions to flavor-violating processes, \emph{e.g.}, $B\to K a$ or $K \to \pi a$. 
These processes are not only induced by flavor-violating ALP couplings, but also arise for flavor-conserving ALP couplings at the one-loop level. In the latter case, the flavor change results from the SM $W$-boson couplings. 
We show two sample diagrams for one-loop contributions to $b \to s a$ transitions for an ALP with flavor-conserving interactions in~\cref{fig:feyn_Ktopi}.\footnote{The explicit calculation of this decay channel can be found, for instance, in this PhD thesis~\cite{Ertas:2021mkt}.}
Depending on the ALP mass and the kinematically open decay channels, different experimental searches focus on the ALP leaving no signature in the detector, e.g.~$K \to \pi + \text{inv}$ at NA62~\cite{NA62:2021zjw}, or its decay to leptons and hadrons, e.g.~at Belle~II~\cite{Belle-II:2023ueh}.

For constraints on flavor-violating ALP couplings see~\cite{Davidson:1981zd,Wilczek:1982rv,Bauer:2019gfk,Cornella:2019uxs,Endo:2020mev,Iguro:2020rby,Calibbi:2020jvd,Davoudiasl:2021mjy,Carmona:2021seb,Cui:2021dkr,Davoudiasl:2021haa,Bauer:2021mvw,Cheung:2021mol,Calibbi:2022izs,Batell:2024cdl,Li:2024thq,Li:2025ski}
for lepton-flavor violating ALPs, and~\cite{Carmona:2022jid,Cheung:2024qve} for ALPs contributing to flavor-violating top decays.

\section{Decay and lifetime}
\label{sec:decay}

Collider search strategies for ALPs depend on its decay channels and lifetime. 
For ALP masses below the di--electron-mass threshold, only the decay $a \to \gamma \gamma$ is possible. 
However, for ALPs with masses in the MeV-GeV range, the ALP Lagrangian in~\cref{eq:ALP_lag} opens up a wide spectrum of kinematically allowed two-body decays into fermions and gauge bosons. 
In the following, we will list the decay width of individual ALP decay channels written in terms of the effective couplings $C_x^\mathrm{eff}$ of the ALP defined in~\cref{eq:Ceff}, which absorb potential loop contributions into their definition.

When considering the decays of the ALP, we need to distinguish the regimes where the QCD is perturbative, $m_a \gg \Lambda_\text{QCD}$, from the region where it is not, $m_a \ll \Lambda_\text{QCD}$, and we need to consider chiral perturbation theory instead. 
We will discuss the different decay channels of the ALP in the following, starting from the perturbative region, $m_a \gtrsim 3$~GeV.

\subsection{Perturbative regime, ALP masses above 3~GeV}
\begin{figure}[t]
    \centering
    \includegraphics[width=0.5\linewidth]{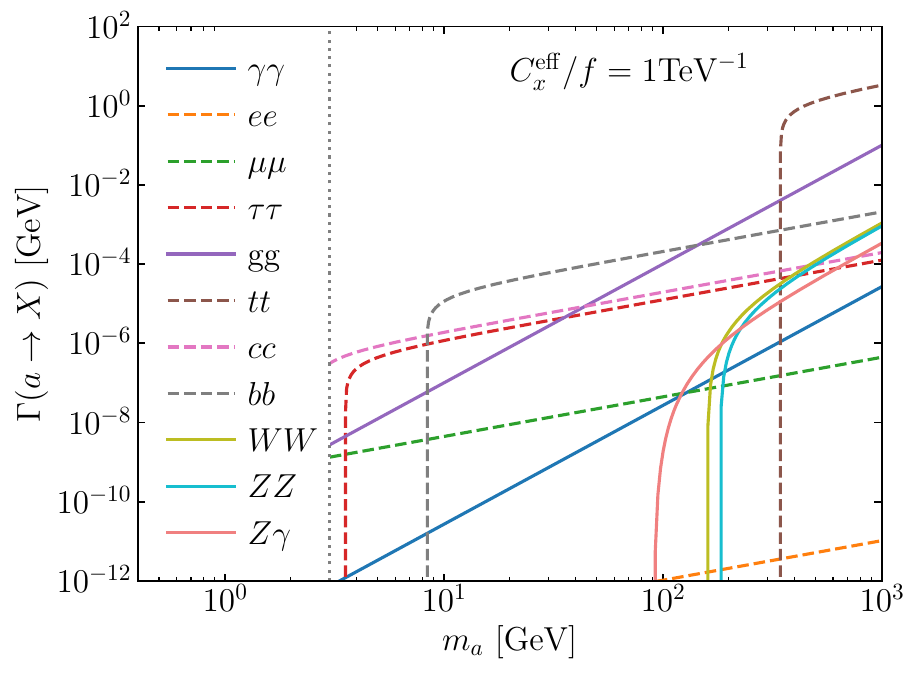}
    \caption{Partial decay widths in individual two-body decay channels of the ALP. 
    Bosonic and fermionic modes are represented by solid and dashed lines, respectively.
    The dotted line at $3$~GeV highlights the beginning of the perturbative regime for the ALP mass. For each mode, a single relevant effective ALP coupling that mediates the decay at tree-level has been switched on with a value of $C^\mathrm{eff}_x/f=1$ TeV$^{-1}$.
    }
    \label{fig:widths_ALP}
\end{figure}
For $m_a \gg \Lambda_\text{QCD}$, the partial widths for kinematically open two-body decays into SM particles are~\cite{Bauer:2017ris,Craig:2018kne}:
\begin{align}
\begin{split}
 \Gamma (a \to \gamma \gamma) &= \frac{\alpha^2 m_a^3}{64 \pi^3 f^2}   \big|\Ceffaa\big|^2 ,  \\
 \Gamma (a \to ff)  &= \frac{m_f^2 \,m_a N_c^f
 }{8 \pi f^2} \big|\Ceffff\big|^2 \sqrt{1 - \frac{4 m_f^2}{m_a^2}} \, ,  \quad f \in \{ e, \mu, \tau, c, b, t \} ,   \\
 \Gamma (a \to gg) &=
 \frac{\alpha_s^2 m_a^3}{8 \pi^3 f^2}  
 \big|\Ceffgg\big|^2
 \left[ 1+ \frac{83}{4} \frac{\alpha_s}{\pi} \right] ,   \\
\Gamma (a \to Z\gamma) &= \frac{1}{32 \pi^3} \frac{m_a^3}{f^2} \frac{\alpha^2 }{\sw^2 \cw^2} \big|\CeffaZ\big|^2 \left(1 - \frac{m_Z^2}{m_a^2} \right)^{3} ,   \\
\Gamma (a \to ZZ) &= \frac{1}{64 \pi^3} \frac{m_a^3}{f^2} \frac{\alpha^2}{\sw^4 \cw^4} \big|\CeffZZ\big|^2 \left(1 - \frac{4 m_Z^2}{m_a^2} \right)^{3/2} ,   \\
\Gamma (a \to WW) &= \frac{1}{32 \pi^3} \frac{m_a^3}{f^2} \frac{\alpha^2}{\sw^4} \big|\CeffWW\big|^2 \left(1 - \frac{4 m_W^2}{m_a^2} \right)^{3/2}  . 
\label{eq:decay_rates}
\end{split}
\end{align}
The decay rate into gluons includes one-loop QCD corrections~\cite{Spira:1995rr,Bauer:2017ris}. 
Note that in addition to the channels listed above, off-shell $VV^*$ decays are relevant in the $[50, 90]$~GeV range~\cite{Craig:2018kne}.
Moreover, loop-effects lead to couplings not present at tree level which can contribute to additional decay channels such as $a \to Z h$~\cite{Bauer:2016ydr,Bauer:2016zfj}. 
In~\cref{fig:widths_ALP}, we show the partial decay rates listed in~\cref{eq:decay_rates} as a function of the ALP mass. Gauge boson decay widths display the characteristic cubic $m_a$ dependence, while fermion decay modes depend linearly on $m_a$ and, like the Higgs boson, are proportional to the square of the fermion mass.
Decays to massive EW gauge bosons are enhanced by inverse powers of $\sw$ with respect to the photon decay rate.
For high ALP masses, decays to gluons and the more massive quarks dominate (for equal effective couplings). The relative size of the contributions obviously depends on the normalization of the Wilson coefficients, especially the fact that the effective couplings to gauge bosons include loop normalization factors $\alpha_i/(4 \pi)$, where $\alpha_i = \alpha, \alpha_s$.

\subsection{Non-perturbative regime, ALP masses below 1~GeV}
\label{sec:decay_non_perturbative}
If the ALP mass is not in the perturbative regime, i.e.\ $m_a \lesssim 1$~GeV, hadronic decay rates and loop contributions need to be calculated using the  chiral effective Lagrangian. 
We do not show the matching of the Lagrangian in~\cref{eq:LEW_mass} onto the chiral effective Lagrangian explicitly here, and refer instead to Section~2.4 of~\cite{Bauer:2021mvw} and references therein.

We briefly discuss hadronic decay modes for ALP masses in the non-perturbative regime: 
The hadronic two-body decays $a \to \pi \pi$ and $a \to \pi^0 \gamma$ are forbidden by \CP\ invariance and angular momentum conservation.
Three-body decay modes involving hadrons and photons or electrons, i.e.\ $a \to \pi \pi \gamma$, $a \to \pi^0 \gamma \gamma$ and $a \to \pi^0 e^+ e^-$, are strongly suppressed by phase space as well as powers of $\alpha$. 
Therefore, the dominant hadronic ALP decay modes below 1~GeV are $a \to 3\pi^0$ and $a \to \pi^+ \pi^- \pi^0$. 
The decay width for these channels is given by 
\begin{align}
    \Gamma (a \to \pi^a \pi^b \pi^0) &= \frac{m_a \,m_\pi^4}{6144 \pi^3 f^2 f_\pi^2} (\Delta c_{ud})^2 g_{ab} \left( \frac{m_\pi^2}{m_a^2} \right) \, ,
\end{align}
where the functions $g_{ab}$ are given for neutral and charged pions as 
\begin{equation}
\begin{split}
     g_{00} (r) &= \frac{2}{(1-r)^2} \int_{4r}^{(1-\sqrt{r})^2} \text{d}z \sqrt{1 - \frac{4r}{z}} \lambda^{1/2} (z,r) \, , \\
     g_{+-} (r) &= \frac{12}{(1-r)^2} \int_{4r}^{(1-\sqrt{r})^2} \text{d}z \sqrt{1 - \frac{4r}{z}} (z-r)^2 \lambda^{1/2} (z,r) \, ,
\end{split} 
\end{equation}
with $\lambda$ defined below \cref{eq:XXp_to_a} and $0\leq r \leq 1/9$ to ensure that the decay is kinematically accessible. 
$\Delta c_{ud}$ is defined as 
\begin{align}
    \Delta c_{ud} \equiv c_{uu} - c_{dd} + 2 \cGG \frac{m_d - m_u}{m_d+m_u} \, .
\end{align}
Finally, it is worth noting that there is a sizable gluon-induced contribution to the effective ALP-photon coupling in the chiral effective Lagrangian.

\subsection{Decay length}
\begin{figure}
    \centering
    \includegraphics[width=0.5\linewidth]{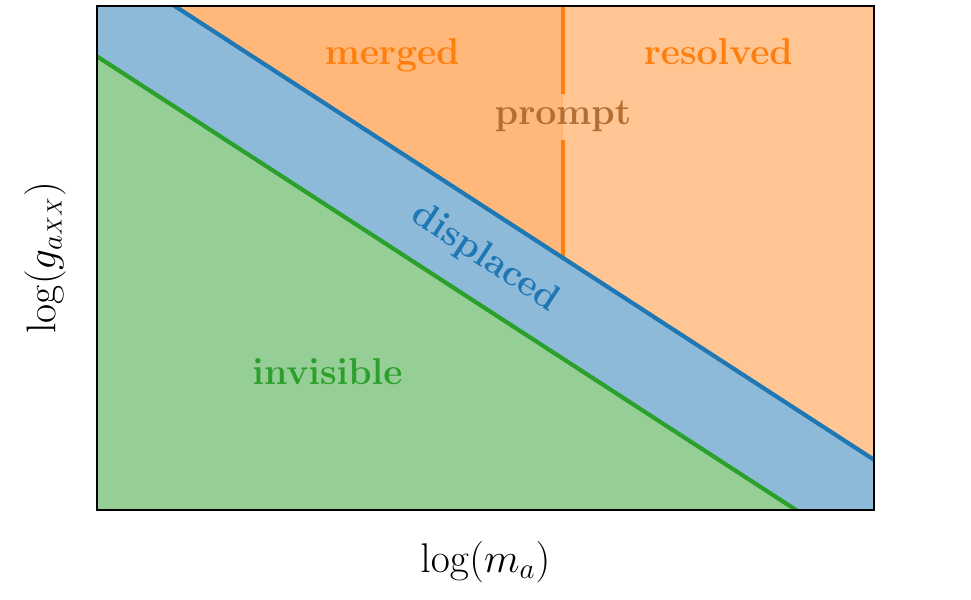}
    \caption{Different kinematic regimes relevant for ALP decays, adapted from~\cite{Dolan:2017osp}.}
    \label{fig:decay_length}
\end{figure}
Experimental signatures of ALPs also rely on the ALP decay length in relation to the size of the detector. 
The ALP decay length~$l_a$ determines whether the ALP decays to SM particles before, within or outside of a particle detector, leaving a prompt, displaced or missing-energy signature, respectively. It can be expressed as 
\begin{align}
l_a &= |\vec{p}_{a, \text{lab}}|/(m_a \Gamma^\text{tot}) \, ,
\end{align}
where $\vec{p}_{a, \text{lab}}$ denotes the ALP momentum in the lab frame. 
The probability density of an ALP decaying after traveling a distance $l$ is given by an exponential function $ \exp (- l/l_a)$. 
For a detector of size $L_\text{det}$, the probability density is 
\begin{equation}
    \text{PDF}_{\text{decay}} (  m_a , \, L_\text{det} ) = 
     \exp \left( - \frac{ m_a  L_\text{det} \Gamma^\text{tot} }{  |\vec{p}_{a,\text{lab}}| } \right) \, .
\label{eq:decay_prob}
\end{equation}
In case a single channel contributes to the decay rate, the decay probability is proportional to $\exp ( g_{aXX}^2 \, m_a^n)$. The $(m_a , \, g_{aXX})$ parameter space can thus be split up into an invisible, displaced and prompt region, as shown in~\cref{fig:decay_length}.
The prompt region can be further split up into a merged and resolved category. 
In the resolved category, both final-state particles can be resolved experimentally. For low ALP masses, the decay products might be so close to each other that they are identified as a single object experimentally. 
As we can see in~\cref{fig:decay_length}, invisible and displaced ALP decay channels are particularly relevant for lower-mass ALPs, for instance for ALPs with $m_a \lesssim 5$~GeV tested in flavor experiments as discussed in \cref{sec:prod_in_decays}.

In practice, we are typically only interested in whether the ALP decays before or after traveling a certain distance in a given direction, \emph{e.g.}\ perpendicular to the beam pipe, such that only the transverse momentum $p_T = |\vec{p}_{a,\text{lab}}| \sin \theta$, where $\theta$ denotes the angle between the ALP direction and the beam pipe, will enter \cref{eq:decay_prob}. 
In that case, we integrate over the angle $\theta$ to obtain the decay probability
\begin{equation}
    \text{prob}_{\text{decay}} (  m_a , \, L_\text{det} ) = 
    \int_0^{\pi/2}
    \text{d}\theta \sin \theta
     \exp \left( - \frac{ m_a  R_\text{det} \Gamma^\text{tot} }{  |\vec{p}_{a,\text{lab}}| \sin \theta } \right) \, ,
\end{equation}
where we used $R_\text{det}$ for the detector length to highlight that we are interested in the radius of the detector. 
For ALPs produced in decays $A \to B \,a$, the ALP momentum in the lab frame is a function of the mass of the ALP as well as the those of the other particles involved in the decay.

\section{Non-resonant and indirect searches}
\label{sec:indirect}
So far, we have discussed on-shell production and decay as a collider probe of ALPs. An alternative and complementary method is to search for indirect signs of these particles through measurements of processes without the ALP in the final state. In other words, the parameters of the ALP-EFT can lead to modifications of processes involving only SM particles, which can be probed by precision measurements. 
\subsection{Non-resonant/off-shell effects}
\label{sec:offshell}
\begin{figure}[h!]
    \centering    \includegraphics[width=0.26\linewidth]{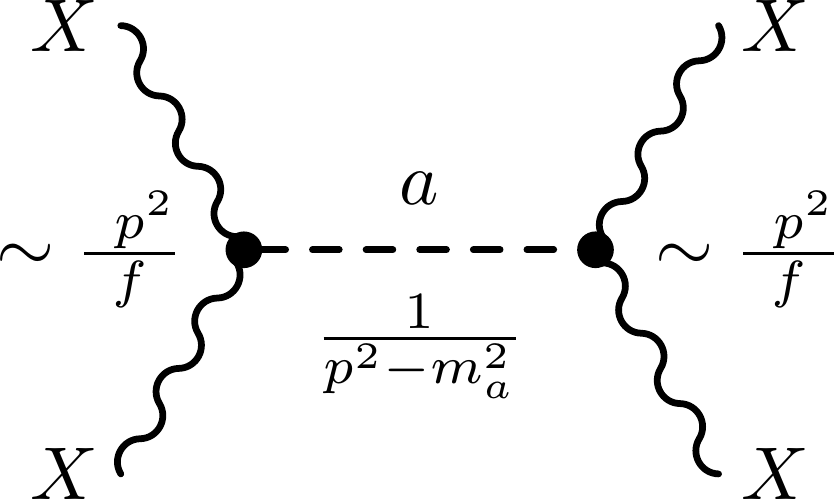}
    \caption{
    \label{fig:offshell}
    Schematic Feynman diagram for the $2\to2$ scattering of SM gauge bosons mediated by an ALP with momentum-dependent couplings.
    }
\end{figure}

 ALPs can manifest themselves indirectly by mediating SM amplitudes non-resonantly or off-shell, at center-of-mass energies $\sqrt{s}>m_a$. As shown in~\cref{eq:gaXXp_FR}, the gauge boson interactions of the ALP scale with the momentum, which leads to amplitudes that grow with energy. This is schematically shown in~\cref{fig:offshell}, representing a Feynman diagram for the ALP-mediated $2\to2$ scattering of SM gauge bosons. 
When $p^2\gg m_a^2,m_X^2$, the momentum-dependent couplings over-compensate the $1/p^2$ suppression from the ALP propagator, leading to an amplitude that grows like $p^2/f^2$. On dimensional grounds, this is the maximal scaling achievable from dimension-five interactions.

Note that, in the off-shell case, the two powers of ALP couplings lead to a $1/f^2$ suppression, such that the effect on the amplitude is at dimension six in the EFT-expansion and the amplitude squared grows with the fourth power of the energy. The cross-section therefore grows with the square of the energy in the high-energy limit. For example, for $s\gg m_a^2$ the unpolarized cross-section for ALP-mediated light-by-light scattering, $\gamma\gamma\to a^\ast \to \gamma\gamma$, is given by
\begin{align}
\sigma(\gamma\gamma\to a^\ast \to \gamma\gamma) = \frac{5\gaa^2s}{24\pi f^4}.
\end{align}
Since the ALP-fermion couplings are proportional to fermion masses (see~\cref{eq:Lferm_eom}), amplitudes involving two fermions and two gauge bosons will instead scale with $(m_f p)/f^2$, and four-fermion amplitudes will tend to a constant $\sim m_f^2/f^2$, like for a Higgs-boson mediated amplitude. 

Overall, the effective couplings of the ALP lead to energy-enhanced off-shell contributions in amplitudes involving gauge bosons. 
These can be probed in high-energy tails of differential distributions of precision SM cross-section measurements. 
The energy growth helps the ALP contributions to be distinguished from typical SM backgrounds, much like indirect new physics searches in the SMEFT framework. 
Several phenomenological studies have been performed along these lines, focusing on LHC searches in diboson final-states, either from gluon-fusion production $gg\to XX^\prime$ or 
in vector-boson-scattering topologies, $q\bar{q}\to XX^\prime q\bar{q}$, as well as $t\bar{t}$ final states \cite{Gavela:2019cmq,Carra:2021ycg,Bonilla:2022pxu,Biswas:2023ksj}. The additional $f^2$ suppression leads to a weaker expected sensitivity than on-shell searches as well as these processes being sensitive to products of ALP couplings. However, the off-shell probes have the benefit of being independent of $m_a$ provided that the processes are measured in the high-energy regime, where the sensitivity is maximized. Moreover, they are also independent of the decay width and branching fractions of the ALP, meaning that they are a  more direct probe of the relevant couplings compared to on-shell production and decay, which depends on all of the ALP couplings through the total width. 

\begin{center}
\begin{figure}[t!]
    \centering
\includegraphics[width=0.26\linewidth]{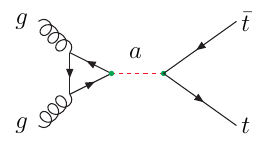}\hspace{0.5cm}
    \raisebox{3pt}{\includegraphics[width=0.26\linewidth]{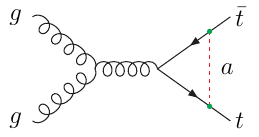}}\hspace{0.5cm}
    \raisebox{2pt}{\includegraphics[width=0.26\linewidth]{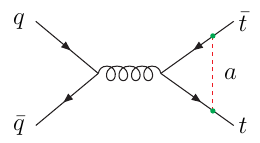}}
    \caption{
    \label{fig:ttbar}
    Selected Feynman diagrams for virtual ALP contributions to $pp\to t\bar{t}$ mediated by the ALP-top coupling, $c_{tt}$, taken from~\cite{Blasi:2023hvb}.
    }
\end{figure}    
\end{center}

ALPs can also impact SM amplitudes indirectly through loop diagrams with virtual ALPs. Although these come with an added loop-factor, the $1/f^2$ suppression is the same as the tree-level, off-shell effects discussed so far. Sufficiently precise measurement can nevertheless lead to sensitivity to couplings that are otherwise difficult to probe, such as the ALP coupling to the top quark, $c_{tt}$. This coupling leads to modifications of differential distributions in top pair production, $pp\to t\bar{t}$, for which a large amount of precise differential LHC data exists.~\Cref{fig:ttbar} shows a selection of relevant Feynman diagrams for virtual ALP corrections to this process.
This top-philic ALP scenario has been studied in several works ~\cite{Ebadi:2019gij,Esser:2023fdo,Blasi:2023hvb,Phan:2023dqw} finding that the indirect loop effects in $t\bar{t}$ data offer competitive bounds to resonant searches in the intermediate mass window, $65<m_a<350$ GeV. Note that the first diagram in~\cref{fig:ttbar} corresponds to a non-resonant ALP contribution to this process, yielding a positive-definite effect which will come to dominate at high invariant masses. 
The virtual contributions, instead, are negative and maximized near the threshold region such that there is an interplay between these two effects across the energy ranges probed by the LHC experiment. 
A non-zero ALP-gluon coupling would, instead, lead to a tree-level contribution to the off-shell topology. 
For similar studies of virtual ALP corrections to Higgs decays and EW precision observables, we refer the reader to Refs.~\cite{Aiko:2023nox}~and~\cite{Aiko:2023trb}, respectively.
\subsection{ALP-SMEFT interference}
As discussed in the introduction, the complete low-energy EFT that extends the SM with an ALP should include operators of the SMEFT, made up only of the SM degrees of freedom. At dimension five, there is only one SMEFT operator consistent with the symmetries and matter content, the so-called Weinberg operator, which violates lepton number by two units and has the schematic form, $(\tilde{H}L)(\tilde{H}\bar{L}^c)$. A large number of possible operators exist at the following, dimension-six order, which are the subject of a wide-ranging program of precision tests of the SM in high-energy physics. 

We have seen that loop contributions from virtual ALPs start to affect SM amplitudes at the order of $1/f^2$ in the EFT expansion, corresponding to dimension six. 
We therefore expect that under RG evolution the ALP couplings should run and mix not only among themselves, as discussed in~\cref{{sec:ALP-RGE}}, but also with the SMEFT Wilson coefficients of dimension six and higher. 
More explicitly, loop diagrams proportional to quadratic forms in the ALP couplings $\sim c_{i}c_{j}/f^2$ generate UV divergences in scattering amplitudes between SM states. These require local counterterms corresponding to SMEFT operators to absorb them and ultimately lead to a contribution to the running of SMEFT Wilson coefficients. 
Schematically, the RG equation of a dimension-six Wilson coefficient, $C^{(6)}_i$, will receive contributions from other dimension-six operators encoded in the well known SMEFT anomalous dimension matrix, $\gamma_{ij}$, as well as from dimension-five ALP interactions, $c_{i}^{(5)}$, through a new anomalous dimension matrix, $\gamma^{ALP}_{ikl}$, as follows
\begin{align}
    \label{eq:ALPSMEFT-RGE}
    \beta_i\equiv\frac{d}{d\ln\mu} C^{(6)}_i(\mu)= \gamma_{ji}(\mu)\,C^{(6)}_{j}(\mu) +  \gamma^{\sss \mathrm{SMEFT-ALP}}_{ikl}(\mu)\frac{c^{(5)}_k(\mu)\, [c^{(5)}_l(\mu)]^*}{(4 \pi f)^2},
\end{align}
where the SMEFT Wilson coefficients are implicitly suppressed by two powers of the cut-off parameter $\Lambda$ and the dependence of the $\gamma$s on the renormalization scale $\mu$ appears through the running of SM parameters. This equation encodes the scale dependence of the dimension-six SMEFT coefficients at scales below $\Lambda$, which is typically assumed to be equal to or somewhat higher than $f$, up to $\Lambda\sim4\pi f$. The exact relation between the scale of the UV theory, $\Lambda$ and $f$ is model-dependent. 

The ALP-EFT contributions to the SMEFT RG evolution, $\gamma^{\sss \mathrm{SMEFT-ALP}}_{ijk}$ were calculated in~\cite{Galda:2021hbr} (see~\cite{Galda:2023qjx} for the ALP-LEFT interference and~\cite{Aebischer:2025zxg} for \CP-violating effects in the ALP-SMEFT interference). In general, the values of the SMEFT coefficients are modified from the high scale down to a lower energy scale at which they may contribute to physical observables. This allows us to place indirect constraints on the ALP-EFT coefficients through precision measurements of SM processes. 
As an example, the coefficient of the dimension-six operator $ \mathcal{O}_{HD}= (H^\dagger D^\mu H)^\ast(H^\dagger D_\mu H)$ 
receives a contribution to its beta function from the ALP coupling to the hypercharge gauge boson in the chirality-flipping basis, $\tcBB$, 
\begin{align}
\label{eq:beta_HD}
    \beta_{HD} \supset \frac{32}{3}\frac{\alpha_1^3}{(4\pi)^3 f^2}\mathcal{Y}_{H}^2\,\tcBB^2,
\end{align}
with $\mathcal{Y}_H=1/2$ being the hypercharge of the Higgs. This operator violates custodial symmetry and is therefore tightly constrained by electroweak precision tests, notably the measurement of the $W$ boson mass, $m_W$.\footnote{In some cases, $m_W$ is used as an input parameter that fixes the remaining parameters of the EW theory. In this event, corrections to $m_W$ are propagated to other predictions of the model, \emph{e.g.} by correcting the relationship between the EM fine structure constant and the EW input parameters. Regardless of the input scheme used, these modifications of the EW sector can be constrained by some combination of precision measurements~\cite{Brivio:2017bnu,Biekotter:2023xle,Celada:2024cxw,Mildner:2024wbl}.} We can use this measurement to infer constraints on the ALP-EFT under certain assumptions.

It can be shown that the relative shift in $m_W$ induced by a non-zero $C_{HD}$ is given by
\begin{align}
    \frac{\delta m_W}{m_W}\sim -0.36 v^2 C_{HD} \, .
\end{align}
We can approximately determine the value of $C_{HD}$ at the scale $\mu=m_W$ by solving the $\beta$-function in~\cref{eq:beta_HD} at the leading logarithmic order, \emph{i.e.} by neglecting the scale dependence of the right-hand side. In terms of the value of $\tcBB$ at some high scale, say $\mu_0=2$ TeV, assuming that all other couplings (including $C_{HD}$ itself) vanish at $\mu_0$,
\begin{align}
    C_{HD} = \beta_{HD}\log\frac{\mu_0}{m_W} =\frac{8 \alpha_1^3}{3(4\pi)^3} \frac{\tcBB^2}{f^2} \log\left(\frac{2\,\mathrm{TeV}}{m_W}\right).
\end{align}
An illustrative relative precision on $m_W$ of $1\times10^{-4}$, implies a bound of $\tcBB/f\lesssim 1$ GeV$^{-1}$.
This is similar to the bound found in~\cite{Biekotter:2023mpd}, where the normalization of the coupling differs by a factor $\alpha_1/(4\pi)$. 

In Ref.~\cite{Biekotter:2023mpd}, a comprehensive, global SMEFT fit to collider data was interpreted in the context of the ALP-EFT, via ALP-SMEFT interference. In many cases, the bounds obtained were found to be competitive or stronger than existing  limits on ALP couplings. Since the sensitivity in these types of analysis arise from the mixing with SMEFT operators, one of the key assumption is that the dominant operators generated at the high scale are those involving the ALP, rather than the general dimension-six SMEFT operators, which are assumed to vanish at the matching scale. The validity of this assumption is model dependent, but it was shown therein that for two explicit UV models the analysis led to conservative constraints on the parameters. A global interpretation in the ALP-SMEFT, taking into account ALP-SMEFT interference, off-shell and finite virtual corrections from ALPs,and direct searches would represent the most general, model-independent approach.

\section{Summary of collider bounds on $\gaa$}
\label{sec:bounds}
Collider bounds on ALPs are an active and fast-evolving field of research. 
A living summary of bounds on ALPs at colliders and beyond is provided in~\cite{AxionLimits}, although we note that at the time of writing it is mostly focused on bounds from astrophysical and cosmological sources.
Here, we discuss collider bounds on the ALP-photon coupling as a well-documented example and highlight some general features and caveats that could apply the study of other ALP couplings.
\begin{figure}
    \centering
    \includegraphics[width=0.75\textwidth]{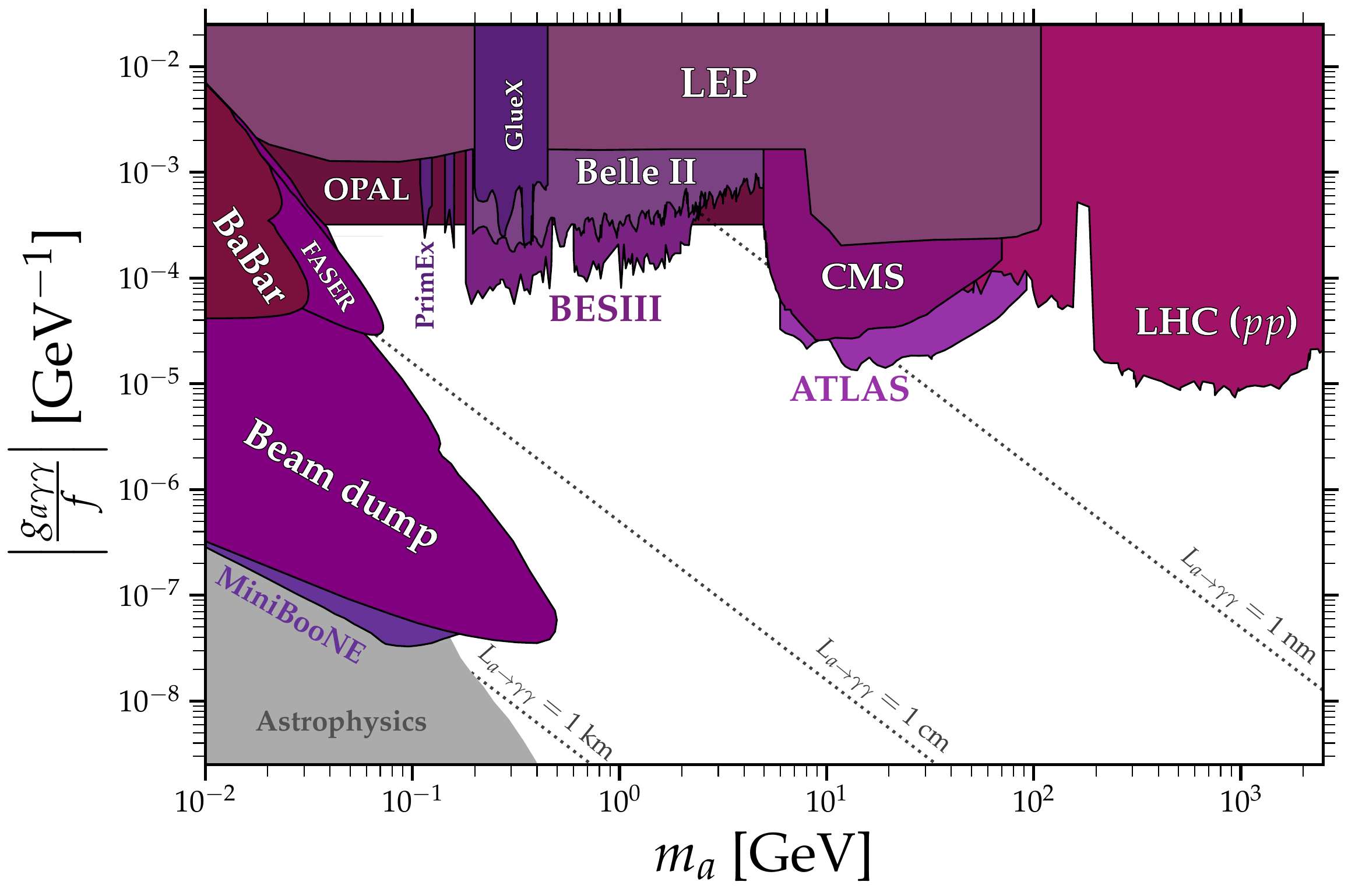}
    \caption{\label{fig:AxionLimits_Zoom}
    Bounds on the ALP-photon couplings shown in~\cref{fig:AxionLimits}, zooming into the high mass region and focusing on collider experiments, adapted from Ref.~\cite{AxionLimits}; primary sources can be found in~\cite{FASER:2024bbl,BaBar:2008aby,BaBar:2010eww,Hearty:1989pq,Riordan:1987aw,Dolan:2017osp,Bjorken:1988as,Dobrich:2019dxc,Blumlein:1990ay,NA64:2020qwq,Capozzi:2023ffu,Jaeckel:2015jla,L3:1995nbq,Belle-II:2020jti,Knapen:2016moh,OPAL:2002vhf,BESIII:2022rzz,BESIII:2024hdv,PrimEx:2010fvg,Aloni:2019ruo,ATLAS:2020hii,CMS:2018erd,ATLAS:2023zfc,TOTEM:2023ewz,Pybus:2023yex}.
}
\end{figure}

In \cref{fig:AxionLimits_Zoom}, we show a zoomed-in version of the plot in \cref{fig:AxionLimits} focusing on the mass range above $10$~MeV, where collider bounds become relevant. 
Note that the limits presented assume an ALP-photon coupling only, with all other ALP couplings set to zero. 
Observables focusing on prompt ALP decays (\emph{e.g.}\ OPAL, LEP, Belle~II, LHC) set a lower bound on the ALP coupling for a given ALP mass, while those sensitive to a specific ALP lifetime (Faser~\cite{FASER:2024bbl}, BaBar~\cite{BaBar:2008aby,BaBar:2010eww,Hearty:1989pq}, beam dumps like CHARM, E141, E137, NuCual, NA64~\cite{Riordan:1987aw,Dolan:2017osp,Bjorken:1988as,Dobrich:2019dxc,Blumlein:1990ay,NA64:2020qwq} and MiniBooNE~\cite{Capozzi:2023ffu}) have a characteristic wedge shape in the mass-coupling plane.
For a given value of $m_a$, the latter observables are thus only sensitive over a specific range of the ALP-photon coupling. 
For large $\gaa$ values the ALP decays before reaching the detector and for small $\gaa$ values not enough ALPs are produced or their lifetime becomes so long that most of them decay after the detector.\footnote{Dashed lines in the plot represent lines of constant proper decay length, $L_{a\to\gamma\gamma}$. To obtain the decay length in the frame of the associated beam dump, the Lorentz boost into the detector rest frame needs to be performed. Therefore, the upper limits of the limits of the Faser, BaBar, beam dump and MiniBooNe limits is not parallel to lines of constant $\gaa$. }
If additional ALP couplings, \emph{e.g.}\ to electrons, were present, the lifetime of the ALP would be modified and the upper limit on $\gaa$ would be shifted to lower ALP-photon couplings.

Different observables/experiments focusing on prompt ALP decays are sensitive to different ALP mass ranges. 
For ALP masses up to several GeV, electron-positron colliders set tight constraints using the $e^+ e^- \to a \gamma$ channel with photonic ALP decays. 
While limits from LEP~\cite{Jaeckel:2015jla} in the $m_\pi \leq m_a \leq 10$~MeV range~\cite{L3:1995nbq} and Belle~II~\cite{Belle-II:2020jti} focus on a $3 \gamma$ final state, limits from OPAL~\cite{Knapen:2016moh,OPAL:2002vhf} assume that the two photons from the ALP decay are so boosted that they will be identified as a single photon experimentally, thus focusing on a $2\gamma$ signature. 
BES~III~\cite{BESIII:2022rzz,BESIII:2024hdv} focuses on the resonant $e^+ e^- \to J/\psi a, a \to \gamma \gamma$ channel. This limit would be affected in the presence of an ALP-charm coupling. 
PrimEx~\cite{PrimEx:2010fvg,Aloni:2019ruo} and GlueX~\cite{Pybus:2023yex,Aloni:2019ruo} constrain Primakoff production ($\gamma N\to aN$, where $N$ is a nucleon or atomic nucleus) with subsequent decay to two photons, targeting different photon beam energies and hence being sensitive to different ALP mass ranges.  
The limits labeled ``ATLAS/CMS'' correspond to data obtained in ultra-peripheral heavy ion collisions~\cite{ATLAS:2020hii,CMS:2018erd}, as discussed in~\cref{subsec:prod_gauge}. They provide the strongest sensitivity to the ALP-photon coupling in the $10\lesssim m_a\lesssim100$ GeV range as a result of the enhancement of the photon luminosity by the large electric charge  of the heavy ions. 
Beyond $100$~GeV, the strongest limits on the ALP-photon coupling is set by the LHC in proton-proton collisions, labeled ``LHC (pp)''~\cite{ATLAS:2023zfc,TOTEM:2023ewz}. 

The sensitivities of all the searches in this mass-coupling plane that rely on an ALP decaying to a pair of photons would be significantly modified in the presence of additional ALP couplings. Such couplings would, in general, allow for additional ALP decays modes and therefore modify the $a\to\gamma\gamma$ branching fraction, or its lifetime. Since the ALP-photon coupling is typically loop-generated and therefore suppressed by $\alpha_{\mathrm{EM}}$, it is likely that the bounds in~\cref{fig:AxionLimits_Zoom} would not directly apply to typical ALP models. A non-zero coupling to gluons of comparable size, for example, would immediately suppress the diphoton branching fraction by a factor of 8 due to color multiplicity as well as a relative factor of $(\alpha_{S}/\alpha_\mathrm{EM})^2\sim 190$, and strongly motivate resonance searches in hadronic final states, instead.
Similarly, as discussed in~\cref{eq:coup_gauge}, the ALP-photon coupling relies on a non-zero coupling of the ALP to the $W$ and $B$ field strength, $\cBB$ or $\cWW$. 
Depending on which of these couplings (or their combination) is responsible for the non-zero value of $\gaa$, additional constraints on the ALP are present. In the mass range below $5$~GeV, strong constraints from flavor physics apply for $\cWW$ (see~\cref{sec:prod_in_decays}). 
For masses above $m_Z$ and $2 m_Z$, the decay channels $a \to Z \gamma$ and/or $a \to ZZ$ open up and modify the ALP branching ratio. This would again modify the LHC limits shown in~\cref{fig:AxionLimits_Zoom}. 
While~\cref{fig:AxionLimits_Zoom} provides a useful first overview of current constraints on the ALP-photon coupling, it is clear that a global analysis (see, \emph{e.g.},~\cite{Bruggisser:2023npd}) including multiple ALP Wilson coefficients is needed to obtain a full picture of the status of constraints and consequently draw more robust conclusions on our sensitivity to ALPs at colliders. 

\section{Conclusions\label{sec:conclusions}}
Axion-like particles are  theoretically well-motivated and arise in various extensions of the SM. Constraining the properties of these BSM particles is an active and growing field of research. From a bottom-up perspective, there is no prior on the mass of these states, and they can be searched for via a wide array of experimental techniques ranging from astronomical observations, to tabletop experiments, to high-energy particle colliders. Colliders, in particular, offer the chance to provide the best tests for ALPs with masses greater than $100$~MeV or so. In this chapter, we have reviewed the effective field theory description of ALPs. We have analysed the interactions of ALPs with SM particles at various energy scales and discussed the dominant production and decay modes through which they might be observed at colliders. Moreover, we have reviewed how precision measurements of SM processes can yield indirect constraints on ALPs from non-resonant production, loops of virtual ALPs and the ALP-SMEFT interference. Although the vast majority of constraints on ALPs focus on one coupling at a time, it is clear that a more global approach is warranted.
We hope that this chapter can serve as a starting point for researchers beginning to work on the collider phenomenology of ALPs, providing them with a solid theoretical foundation as well as a source of references to refine their knowledge. Ref.~\cite{Tammaro:2025zso}, for example, provides an introduction to searches for dark sectors at colliders with an emphasis on ALPs, focusing more on B-factories and meson decays. For those who are interested in ALPs beyond the collider context, \emph{e.g.} in astrophysics, cosmology and low-energy experiments, we refer to~\cite{Graham:2015ouw,Irastorza:2018dyq,Choi:2020rgn,Galanti:2022ijh}. For a pedagogical introduction to axions and the strong \CP\ problem, see~\cite{Marsh:2017hbv,Yu:2023gdq}.

\section*{Acknowledgements}

We would like to thank Martin Bauer, Ilaria Brivio, Simone Blasi, Claudia Cornella, and Simone Tentori for useful discussions. 
The research of A.B.\ is supported by the Deutsche Forschungsgemeinschaft
(DFG, German Research Foundation) under grant 396021762 - TRR 257.
K.M.\ is supported by an Ernest Rutherford Fellowship from the STFC, Grant No.\ ST/X004155/1.
We would also like to express special thanks to the Mainz Institute for Theoretical Physics (MITP) of the Cluster of Excellence PRISMA+ (Project ID 390831469), for its hospitality and its partial support during the completion of this work.

\newpage

\bibliographystyle{Numbered-Style}
\bibliography{reference}

\end{document}